\documentclass{IEEEtran}
\usepackage[hyphens]{url}
\usepackage{enumerate}
\usepackage{graphicx}
\usepackage{multirow}
\usepackage{multirow, enumitem, subfigure}
\usepackage[usenames,dvipsnames]{xcolor}
\usepackage{amsmath}
\usepackage{mathtools}
\usepackage{tikz}
\usepackage[normalem]{ulem}
\usepackage{caption}
\usepackage{color}
\usepackage{soul, textcomp}
\usepackage{comment}
\usepackage{flushend}

\begin{document}

\title{\textbf{\LARGe Domain-Sharding for Faster HTTP/2 in Lossy Cellular Networks\vspace{-8pt}}}

\author{ 
    \IEEEauthorblockN{\Large{Utkarsh Goel}\IEEEauthorrefmark{1}, 
    \Large{Moritz Steiner}\IEEEauthorrefmark{1}, 
     \Large{Mike P. Wittie}\IEEEauthorrefmark{2},
    \Large{Stephen Ludin}\IEEEauthorrefmark{1},
     \Large{Martin Flack}\IEEEauthorrefmark{1}
    }\\
    \vspace{1pt}
    \IEEEauthorblockA{\IEEEauthorrefmark{1}\large{Akamai Technologies, Inc. -- \textit{\{ugoel, moritz, mflack, sludin\}@akamai.com}}}\\
    \IEEEauthorblockA{\IEEEauthorrefmark{2}\large{Montana State University -- \textit{mwittie@cs.montana.edu}}}
}

\maketitle
\pagestyle{empty}
\thispagestyle{empty}

\begin{abstract}

HTTP/2 (\texttt{h2}) is a new standard for Web communications that already delivers a large share of Web traffic. 
Unlike HTTP/1, \texttt{h2} uses only one underlying TCP connection. 
In a cellular network with high loss and sudden spikes in latency, which the TCP stack might interpret as loss, using a single TCP connection can negatively impact Web performance. 
In this paper, we perform an extensive analysis of real world cellular network traffic and design a testbed to emulate loss characteristics in cellular networks. 
We use the emulated cellular network to measure \texttt{h2} performance in comparison to HTTP/1.1, for webpages synthesized from HTTP Archive repository data. 
\looseness -1

Our results show that, in lossy conditions, \texttt{h2} achieves faster page load times~(PLTs) for webpages with small objects. 
For webpages with large objects, \texttt{h2} degrades the PLT. 
We devise a new domain-sharding technique that isolates large and small object downloads on separate connections.
Using sharding, we show that under lossy cellular conditions, \texttt{h2} over multiple connections improves the PLT compared to \texttt{h2} with one connection and HTTP/1.1 with six connections. 
Finally, we recommend content providers and content delivery networks to apply \texttt{h2}-aware domain-sharding on webpages currently served over \texttt{h2} for improved mobile Web performance.
\looseness -1

\end{abstract}


\vspace{-12pt}
\section{Introduction}
\label{sec:introduction}

The widely adopted HyperText Transfer Protocol~(HTTP/1.1) improves performance of the request-response model by pipelining requests on the server side.
However, with HTTP/1.1~(\texttt{h1}) servers transfer resources only in the order they receive requests.
This behavior of \texttt{h1} creates a head-of-line~(HOL) blocking for all requests on the application layer and slows down the webpage load time~(PLT), especially when faster requests have to wait in the pipeline for the slower requests to finish processing.
\looseness -1

To mitigate this HOL blocking, modern Web browsers and content providers~(CPs) take different approaches.
Specifically, Web browsers establish up to six TCP connections for each domain name on the webpage~\cite{grigorik2013high}, which allows the browser to request six objects in parallel, but wait for the seventh object until one of the six connections becomes available. 
On the other hand, CPs design their websites such that the embedded objects are distributed across multiple domain names, a technique known as Domain-Sharding~\cite{grigorik2013high}.
The combination of the above two techniques allows Web browsers to establish six connections for each of the many domain names to perform parallel downloading of resources and thus mitigate the HOL blocking.
\looseness -1

The HTTP/2~(\texttt{h2}) protocol, standardized in May 2015, follows a different approach to eliminate HOL blocking on the application layer~\cite{http2:rfc}.
Instead of using domain sharding and opening multiple TCP connections for each domain name, \texttt{h2} uses only one TCP connection to exchange all request and response payloads.
Specifically, instead of relying on pipelining all requests as in the case of \texttt{h1}, \texttt{h2} allows multiplexing and interleaving of all requests and responses such that incoming requests could be processed in parallel and in any order determined by the server.
\looseness -1

While \texttt{h2} eliminates HOL blocking on the application layer, it retains HOL blocking on the transport layer~\mbox{\cite{http2:rfc,quic,ludinH2}}.
Specifically, since \texttt{h2} utilizes TCP as its transport protocol, packet loss on \texttt{h2}'s single TCP connection reduces the congestion window by 50\%~(and 30\% in the case of TCP~CUBIC~\cite{cubic}).
While \texttt{h1} also utilizes TCP and suffers similarly in the event of packet loss, for \texttt{h1} Web browsers establish six TCP connections and packet loss does not degrade the cumulative congestion window as much as it does for the single TCP connection in the case of \texttt{h2}.
Previous studies have shown either improvement or degradation in PLTs when loading webpages over \texttt{h2}, compared to \texttt{h1}~\mbox{\cite{Bocchi2017,7179400,Erman:2013:TSM:2535372.2535399,7557456,moritzHttpWorkshop16,pamhttp2,Wang:2014:SS:2616448.2616484}}.
The disagreement about \texttt{h2} performance creates uncertainty among CPs and their surrogate proxy infrastructures~(content delivery networks, CDNs) as to whether and how to follow suit. 
Should CDNs continue to serve mobile Web content over \texttt{h1}, or argue to their CP customers the performance benefits of serving webpages over \texttt{h2} in mobile networks?
\looseness -1

In this paper, we answer these questions through a comprehensive comparison of \texttt{h1} and \texttt{h2} performance in cellular networks. 
Specifically, we investigate the impact of packet loss on PLT when using \texttt{h1} over six TCP connections and \texttt{h2} over one TCP connection. 
We also investigate whether the webpage structure has implications on \texttt{h2} performance.
In order to understand Web performance, we first extend our previous work to understand the dynamic nature of cellular network characteristics in terms of packet loss, round trip time, and bandwidth~\cite{GoelH2MobiCom16}.
Specifically, we develop several techniques to emulate various cellular network conditions as observed for cellular carriers in rural and urban areas.
We then replay the network conditions and compare PLTs using \texttt{h2} and \texttt{h1} as application layer protocols.
\looseness -1

\noindent
We classify our five major contributions as follows:
\looseness -1

\vspace{2pt}
\noindent
\textbf{Dataset Richness:} Our analysis of cellular network characteristics is based on 50\,K TCP connections captured over several days from an Akamai CDN cluster hosted inside a data-center of a major cellular network provider in the US.
Our collected TCP traces represent the characteristics of a real world cellular network, as the TCP connections to the selected Akamai CDN cluster are not influenced by any interference from the public Internet~\cite{aanp}.
This is because the selected CDN cluster serves traffic only to cellular clients in that ISP.
We provide the details on our data collection technique in Section~\ref{sec:data_collection}.
\looseness -1

\vspace{2pt}
\noindent
\textbf{Identifying Cellular Network Characteristics:} Our analysis of live TCP traces captured from a CDN cluster suggests three unique characteristics of cellular networks.
\begin{itemize}[leftmargin=0.15in]
\item We observe that about 32\% of the TCP connections over cellular networks experience packet loss.
We also make similar observations in our previous work of detecting TCP terminating proxies across cellular networks worldwide, where we use the packet loss information in the TCP logs provided by Linux kernel of production Akamai CDN servers~\cite{GoelMiddlebox16}.
\looseness -1

\item Next, we observe that packet losses in cellular networks are clustered -- several consecutive TCP segments are lost at the same time.
Our observations are similar to the recent work of Flach~\textit{et~al.}, where they found that traffic policers introduce high packet loss rate in cellular networks~\cite{45411}.
\looseness -1

\item Finally, we observe that TCP connections over cellular networks experience packet loss multiple times during their lifetime. 
Additionally, when a loss occurs, up to 40\% of the TCP segments flowing over the connection at that time are dropped.
Note that this loss rate is different from the cumulative packet loss the connection experiences during its entire lifetime.
\looseness -1

\end{itemize}

\vspace{0pt}
\noindent
\textbf{Measurements:} Our comprehensive view of cellular network characteristics from an Akamai CDN cluster deployed inside cellular ISP datacenter allows us to improve existing cellular network emulation techniques.
Specifically, existing network emulators, such as TC~NETEM~\cite{tc}, Network Link Conditioner~\cite{linkConditioner}, and others, introduce packet loss on a link at random times during emulation.
An emulation model dependent on random occurrences of loss does not provide a realistic representation of cellular network conditions.
Therefore, based on the observations we make about cellular network characteristics, we emulate several cellular network conditions in terms of packet loss rate, time gap between two loss events, round trip time~(RTT) between the client and the CDN server, and bandwidth attributed to cellular base stations.
To allow further research in this direction, we make our emulation script available at \texttt{\url{https://github.com/akamai/cell-emulation-util}}.
\looseness -1

\vspace{2pt}
\noindent
\textbf{Investigating HTTP/2 Performance:} We make extensive use of the HTTP Archive repository to synthesize several webpages that represent real world websites~\cite{httparchive}.
For example, we synthesize a webpage with hundreds of small objects~(less than 1\,KB each), a webpage with a few large objects~(about 435\,KB each), and several webpages with tens of objects ranging from 1\,KB to 620\,KB in size. 
Using the emulated network, we conduct several experiments to compare PLTs of these webpages loaded in turn over \texttt{h2} and \texttt{h1}.
We list the results from our experimental evaluation as follows:
\looseness -1

\begin{itemize}[leftmargin=0.15in]

\item We observe that for a webpage with hundreds of small sized objects, \texttt{h2} outperforms \texttt{h1} in all emulated network conditions, except in the condition where the single TCP connection of \texttt{h2} frequently experiences loss.
\looseness -1

\item Next, we observe that for a webpage with a few large objects, \texttt{h2} does not outperform \texttt{h1} at all. 
In fact, the PLTs over \texttt{h2} are significantly higher than \texttt{h1}.
\looseness -1

\item And finally, we observe that for a webpage with object sizes ranging from 1\,KB to 620\,KB, \texttt{h2} outperforms \texttt{h1} when the webpage has few large objects and many small objects.
However, as the number of large objects increase~(keeping the total number of objects on the webpage same), the performance of \texttt{h2} is comparable to \texttt{h1}.
Additionally, as the network conditions on \texttt{h2}'s single TCP connection worsen, PLTs observed over \texttt{h2} are much larger than the PLTs observed over \texttt{h1} under the same network conditions.
\looseness -1
\end{itemize}

\vspace{0pt}
\noindent
\textbf{Sharding Webpages for Faster PLTs over HTTP/2:} To the best of our knowledge, there is currently no known best practice as to how \texttt{h2} should be tuned to minimize the impact of loss on PLT.
In this work, we side-step from the recommendation of disabling domain-sharding in \texttt{h2}~\cite{disableSharding1,http2:rfc,grigorik2013high,disableSharding2,disableSharding3,disableSharding4}.
Our goal is not to argue that domain-sharding is necessary to speedup PLTs for \texttt{h2}-enabled webpages -- only future evaluations will demonstrate the relative benefits of domain-sharding and other methods for reducing PLTs.
Instead, we investigate whether a technique that enables application layer control of how Web objects are downloaded
can be effective and safe in its own right.
We now discuss its implications on Web performance as follows: 
\looseness -1

\begin{itemize}[leftmargin=0.15in]
\item We perform experimental evaluation to demonstrate that multiple \texttt{h2} connections improve the mobile Web performance in lossy cellular network conditions, when compared to both \texttt{h2} with single TCP connection and \texttt{h1} with six TCP connections.
\looseness -1

\item We investigate and develop a new domain-sharding technique that isolates large downloads on separate TCP connections, while keeping downloads of small objects on a single connection.
Our devised domain-sharding technique is different from the legacy sharding technique currently used in the case of \texttt{h1}.
Specifically, current sharding techniques distribute webpage resources over several domain names depending upon the type of object. 
For example, the domain \texttt{img.example.com} is used for all images irrespective of the image size. 
\looseness -1

\item Through experimental evaluation, we demonstrate that \texttt{h2} achieves faster or comparable PLTs to \texttt{h1} when using our domain-sharding technique. 
The PLTs over \texttt{h2} when using the legacy sharding technique results in worse PLTs.
Therefore, we recommend CPs and CDNs to apply \texttt{h2}-aware domain-sharding practices for \texttt{h2}-enabled webpages.
\looseness -1

\end{itemize}

\noindent
The rest of the paper is organized as follows. 
In section~\ref{sec:related_work}, we discuss related work that investigates and improves Web performance over \texttt{h2}.
In Section~\ref{sec:why_testbed}, we discuss our need to develop an emulation testbed, instead of using already deployed systems, for analyzing cellular network characteristics.
Next, we discuss our data collection process in Section~\ref{sec:data_collection} and our characterization of cellular network conditions in Section~\ref{sec:setup}.
In Section~\ref{sec:emulation}, we discuss various emulated cellular network conditions.
In Section~\ref{sec:results}, we compare Web performance over \texttt{h1} and \texttt{h2} and present a novel domain-sharding technique to speed up PLTs in Section~\ref{sec:sharding}.
In Section~\ref{sec:validation}, we discuss the challenges in validating our mobile emulator.
In Section~\ref{sec:discussion}, we discuss the practical implications of applying domain-sharding.
Finally, we conclude in Section~\ref{sec:conclusions}.
\looseness -1

\vspace{-7pt}
\section{Related Work}
\label{sec:related_work}
\vspace{-2pt}

While the \texttt{QUIC} protocol is being designed to eliminate HOL blocking on both transport and application layers, \texttt{QUIC} is not a standard yet and will likely require several years before its widespread adoption by CPs for the production traffic~\cite{quic}.
The BBR protocol is another transport layer development that allows faster loss discovery/recovery in high loss scenarios~\cite{bbr}.
However, our results show that domain-sharding improves PLTs across various emulated high and low loss scenarios.
Therefore, we argue that even with a loss-resistant TCP stack, multiple connections will still enable \texttt{h2} to achieve faster PLTs.
\looseness -1

A study by Mi~\textit{et~al.} shows the performance of a modified version of standard \texttt{h2} protocol between custom clients and Web servers that establishes new TCP connections every time the server detects a large object being requested by the client~\cite{smig}.
While the study shows \texttt{h2} performance benefits when isolating large downloads on separate TCP connections, the modified protocol adds latency to the overall PLT by establishing new TCP connections when a request for large object is detected by the server.
In contrast, we provide an application layer solution that minimizes the impact of TCP HOL blocking on \texttt{h2}'s performance, without the need of any changes to the client Web browser. 
Specifically, our work improves PLTs over \texttt{h2} by establishing multiple TCP connections as soon as the domain-sharded HTML for the base page is available and parsed by the Web browser.
\looseness -1

A study by Varvello~\textit{et~al.} investigates the adoption and performance of \texttt{h2} for top ranked webpages, when measured from wired and mobile hosts~\cite{pamhttp2}.
Their evaluation of PLT over \texttt{h2} indicates performance improvement for some websites and degradation for others, due to reasons associated with latency between hosts and Web servers.
Bocchi~\textit{et~al.} also measure \texttt{h2} performance, however they stop short of explaining reasons behind the observed performance differences~\cite{Bocchi2017}. 
Other studies compare performance of \texttt{h2} without TLS support with \texttt{h1} in clear-text, using 3G USB modems~\cite{7179400} and network simulations using static configurations on TC NETEM~\mbox{\cite{7179400,7557456}}.
Emran~\textit{et~al.} compare SPDY performance with \texttt{h1} over a production cellular network~\cite{Erman:2013:TSM:2535372.2535399}.
Their results indicate that SPDY degrades Web performance in lossy cellular conditions.
\looseness -1

Previous studies collectively do not suggest clear improvements in Web performance when using \texttt{h2} protocol for mobile content delivery.
In contrast to these studies, our goal is to disambiguate previous results through controlled experiments using synthetic webpages that represent structures of popular webpages.
Therefore, we investigate PLTs for synthetic webpages and show the impact of object size on PLT, when using \texttt{h2} in dynamically changing lossy cellular network conditions.
\looseness -1

A study by Wang~\textit{et~al.} investigates the impact of SPDY protocol on Web performance in various simulated cellular network conditions~\cite{Wang:2014:SS:2616448.2616484}.
Their results indicate that SPDY helps for some pages but hurts for others.
Additionally, their study implements domain-sharding as it appears on many webpages using \texttt{h1}, i.e. sharding objects by their type. 
The authors then investigate the impact of domain-sharding on PLT and identify that sharding does not  improve Web performance.
In contrast, our work investigates the impact of domain-sharding based on the object size and make a case for applying domain-sharding to download all small objects one connection and large objects on separate connections.
\looseness -1

\vspace{-7pt}
\section{Why a New Cellular Emulation Testbed?}
\label{sec:why_testbed}

Akamai's global infrastructure for content delivery serves webpages for its CP customers and processes a total of over 40\,million HTTP requests every second from clients around the globe.
Many of its CP customers desire to understand the performance of how fast their websites load on clients in different networks.
To understand the performance of Web content delivery on production traffic, Akamai utilizes Real User Measurement~(RUM)~\cite{akamai:rum} system that uses the Web browser exposed Navigation Timing API~\cite{navigationTimingApi} to capture several performance metrics pertaining to webpage load.
Specifically, when clients request the base page HTML, Akamai's RUM system injects custom JavaScript that runs on the client's browser and uses the Navigation Timing API to record the time taken to perform DNS lookup of the base page domain name, time to establish TCP connection, and the overall PLT, among many other metrics~\cite{GoelIPv6,GoelMiddlebox16}.
While RUM allows for large scale measurements across many mobile networks and client devices, it does not record transport layer metrics, such as packet loss rates and timestamps of loss occurrence during the lifetime of a TCP connection.
These problems have also been discussed at the recent PAM conference~\cite{Bischof2017}.
\looseness -1

Moreover, a recent study shows that webpages are loaded over \texttt{h2} only by modern hardware and Web browsers, whereas \texttt{h1} is used predominantly by older hardware and Web browsers~\cite{moritzHttpWorkshop16}.
Given that mobile hardware significantly impacts the PLT~\cite{DBLP:journals/corr/SteinerG16},  RUM   measurement data pertaining to \texttt{h2} and \texttt{h1} performance is skewed.
It is possible to configure CDN clusters to serve webpages over \texttt{h2} only 50\% of the time and thus perform a fair comparison with \texttt{h1} when webpages are loaded over the same mobile hardware. However, modern webpages contain many resources that are often downloaded over transport and application layer protocols different than the ones used by the base page~\cite{GoelThird},~e.g.~the base page being loaded over \texttt{h2} but most other objects over \texttt{h1}.
Using RUM  for such webpages will further skew the PLTs in our analysis of real world performance of \texttt{h2} and \texttt{h1}.
\looseness -1

Finally, existing mobile network measurement tools, such as Akamai Mobitest~\cite{mobitest}, WebPageTest~\cite{webpagetest}, RadioOpt Traffic Monitor~\cite{radioopt}, Gomez Last-Mile~\cite{gomez}, Keynote~\cite{keynote}, 4GMark~\cite{4GMark}, nPerf~\cite{nperf}, and WProf~\cite{wprof} also do not emulate or record the live loss rates found in real world cellular networks when measuring Web performance.
In fact, many of these and other tools rely on Navigation Timing API and inherit its limitations as discussed earlier~\cite{GoelSurveyE2E}.
\looseness -1

Given the aforementioned limitations of existing Web performance measurement APIs/tools, we develop a testbed to emulate different cellular network conditions by investigating live TCP traffic captured from a real world cellular network.
\looseness -1

\begin{figure*}[t]
\centering
\minipage{0.325\textwidth}
  \includegraphics[width=\linewidth]{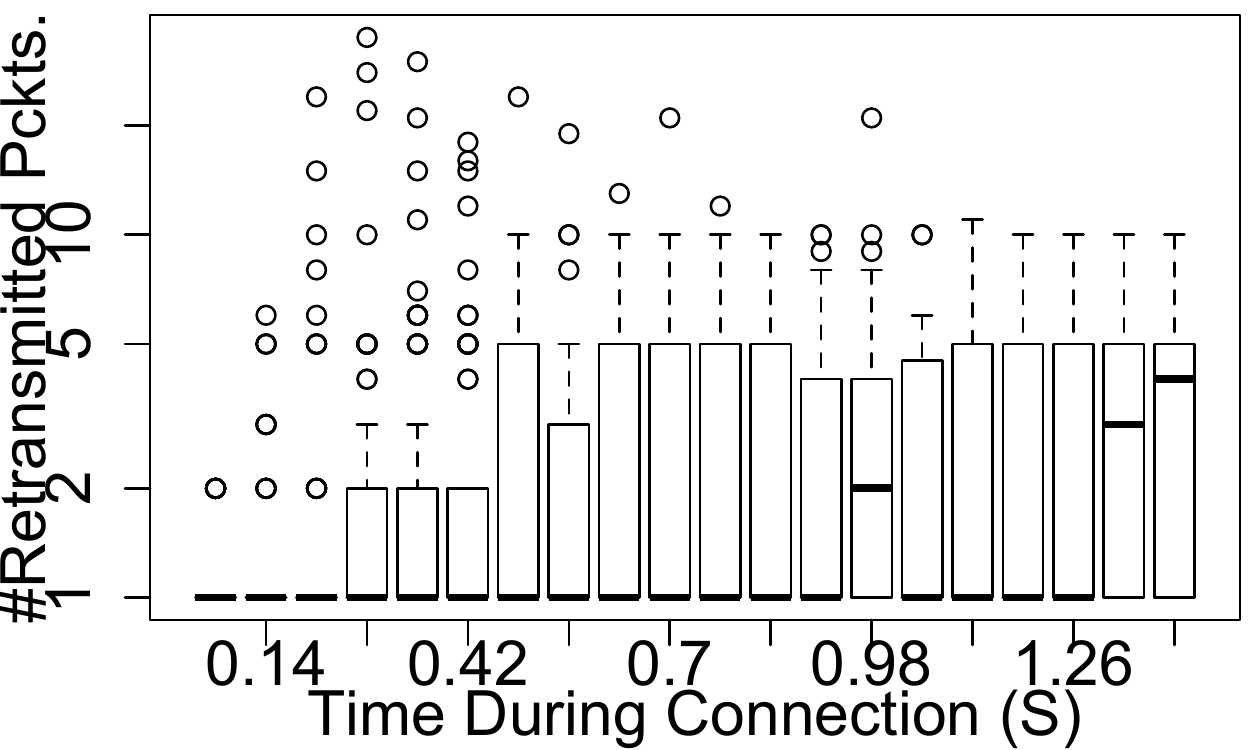}
\vspace{-15pt}
\caption{Clustered retransmissions at different times during connections.}
\label{fig:cluster}
\endminipage
\hfill
\minipage{0.325\textwidth}
  \includegraphics[width=\linewidth]{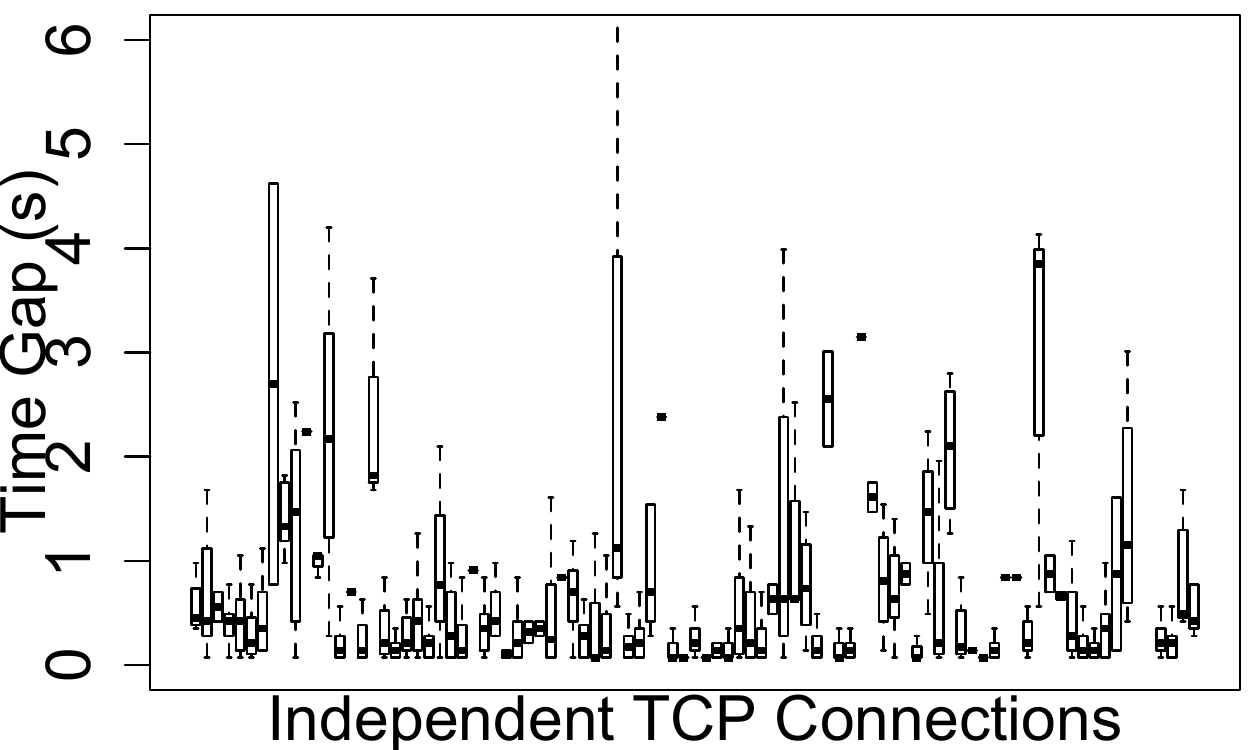}
\vspace{-15pt}
\caption{Time gap between retransmission clusters for different connections.}
\label{fig:gap}
\endminipage
\hfill
\minipage{0.325\textwidth}
  \includegraphics[width=\linewidth]{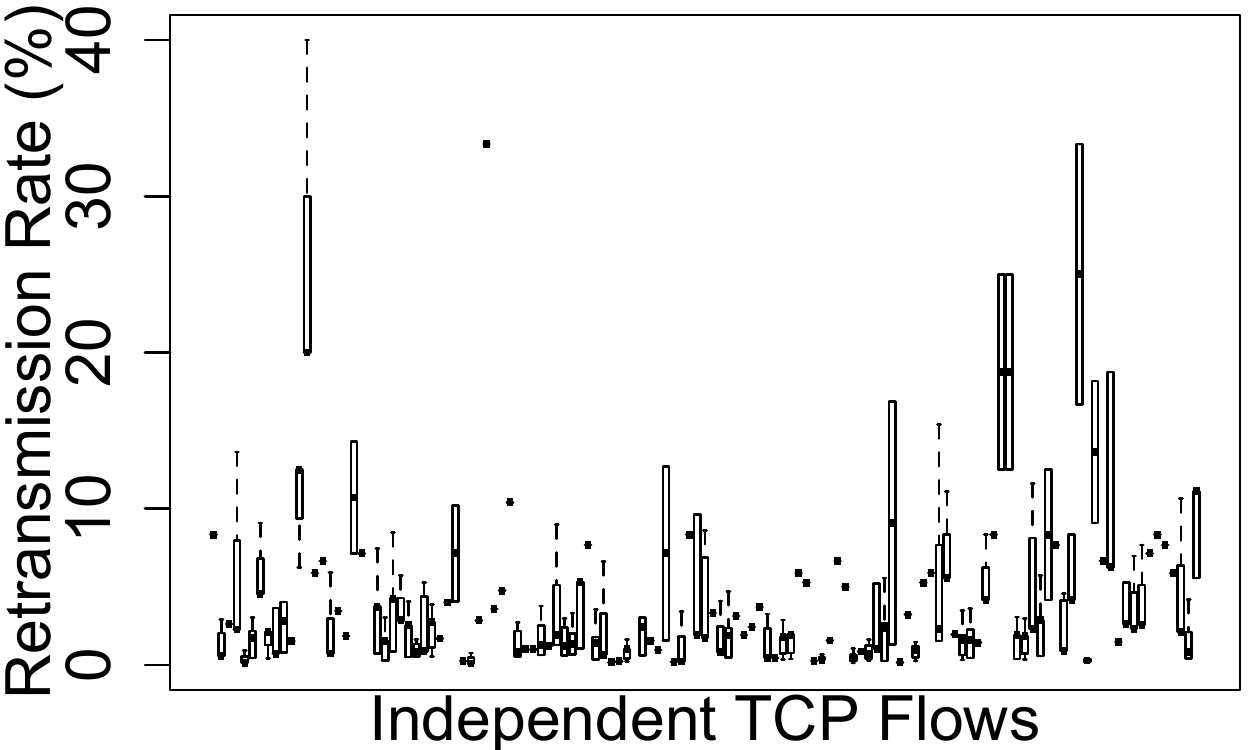}
\vspace{-15pt}
\caption{Retransmission rates for different connections.}
\label{fig:rtx}
\endminipage
\vspace{-15pt}
\end{figure*}

\vspace{-5pt}
\section{Data Collection Methodology}
\label{sec:data_collection}

Akamai accelerates its content delivery by deploying its CDN clusters in close proximity to many cellular ISPs' packet gateways~(P-GWs)~\cite{GoelMiddlebox16,Xu:2011:CDN:1993744.1993777,Zarifis2014}.
In fact, many of Akamai's CDN clusters are deployed deep inside network datacenters of cellular ISPs~\cite{aanp}.
We selected one such CDN cluster co-located with a P-GW of a major cellular ISP in the US
to passively capture live TCP traffic pertaining to that cellular network.
Our choice of the selected CDN cluster is based on the fact that the TCP connections to the cluster are not at all influenced by any interference from the outside public Internet and that the cluster is one of the biggest Akamai clusters hosted in that cellular ISP.
This is because the selected CDN cluster only serves Web content to cellular clients in that ISP~\cite{aanp}.
\looseness -1

Next, we ran \texttt{TCPDump} on each CDN server in the cluster 
at different times of a day to capture incoming and outgoing TCP segments.
Previous works, including our own, show that the cellular network chosen in this study uses TCP terminating proxies to split TCP connections between mobile clients and CDN servers on port 80, but never does so for connections to port 443~\cite{GoelMiddlebox16,ethanCellProxy}.
Therefore, we only capture TCP segments to-and-from port 443, which ensures that the captured segments are for end-to-end connections between clients and CDN servers, as opposed to segments from the TCP terminating proxy~\cite{ethanCellProxy}.
\looseness -1

Next, for each TCP connection captured in \texttt{TCPDump}s, we use \texttt{tshark} to extract four characteristics periodically  every 70\,ms~\cite{tshark}.
The four characteristics are the number of segments exchanged between the client and the server, the number of bytes exchanged between the client and the server, the number of segments retransmitted by the server, and the average time lapse between acknowledgments.
Note that the first 70\,ms interval starts when the \texttt{TCP SYN} is received by the server.
The choice of 70\,ms as the time interval to calculate the above four metrics is based on our previous work showing that 70\,ms is the median Round Trip Time~(RTT) between clients in the chosen cellular network and the selected Akamai CDN cluster~\cite{GoelMiddlebox16}.
Therefore, in the median case we expect the above mentioned three metrics to change after 70\,ms.
\looseness -1

Our total dataset consists of the above mentioned metrics for about 50\,K TCP connections captured on port 443.
Note that the number of connections captured are limited to 50\,K because these connections represent only the HTTPS traffic captured to-and-from one of the P-GWs deployed by the chosen cellular ISP.
We observe that the quality of TCP connections analyzed in this study is consistent with the quality analyzed in previous work conducted at a different time from the same P-GW~\cite{GoelH2MobiCom16}.
Therefore, we argue that our way of profiling TCP connections is independent of the time when TCP traffic is captured.
Further, although in our investigation we capture TCP traffic from one cluster co-located with one P-GW, our analysis of loss from the captured TCP traffic indicates similar packet loss to what is found in many other networks in North America and Europe when analyzing loss from different CDN clusters~\cite{45411,GoelMiddlebox16}.
Therefore, we speculate that our findings in this study are potentially applicable to other cellular networks.
\looseness -1

\begin{figure*}[t]
\centering
\minipage{0.24\textwidth}
  \includegraphics[width=\linewidth]{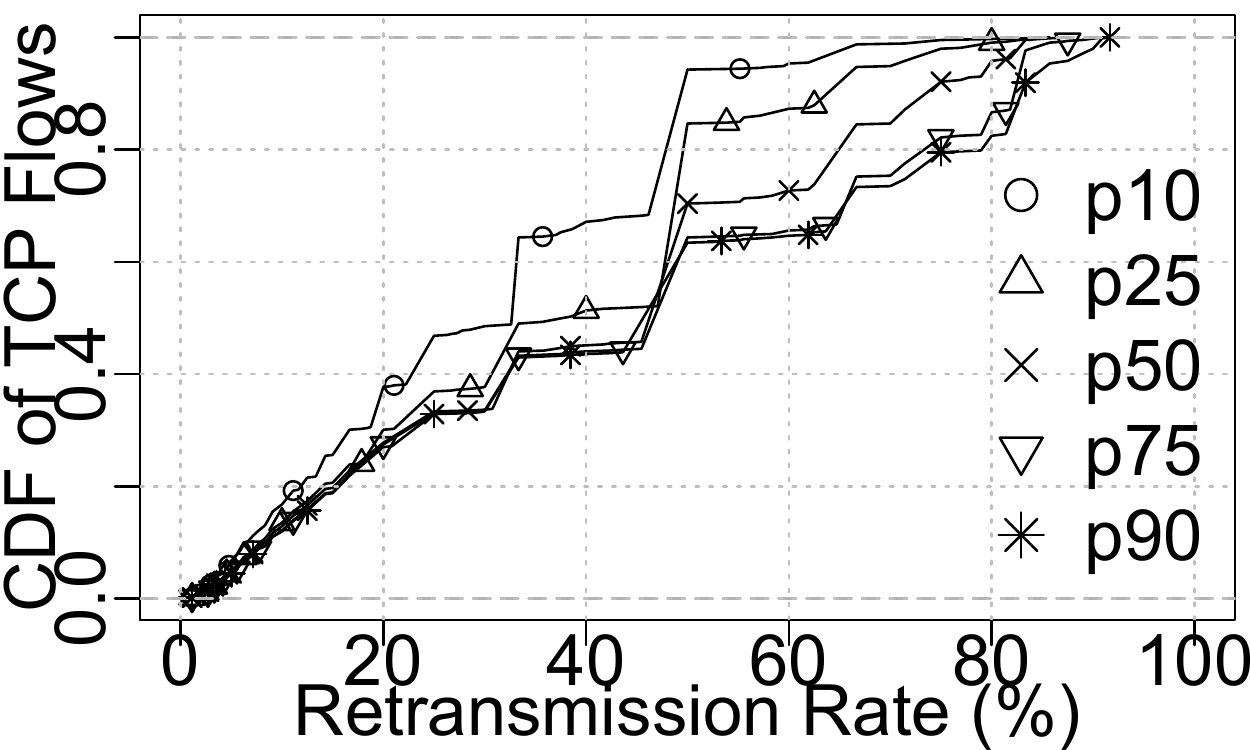}
\vspace{-18pt}
\caption{Distributions of retransmission rates.}
\label{fig:user_dist_rtx}
\endminipage
\hfill
\minipage{0.24\textwidth}
  \includegraphics[width=\linewidth]{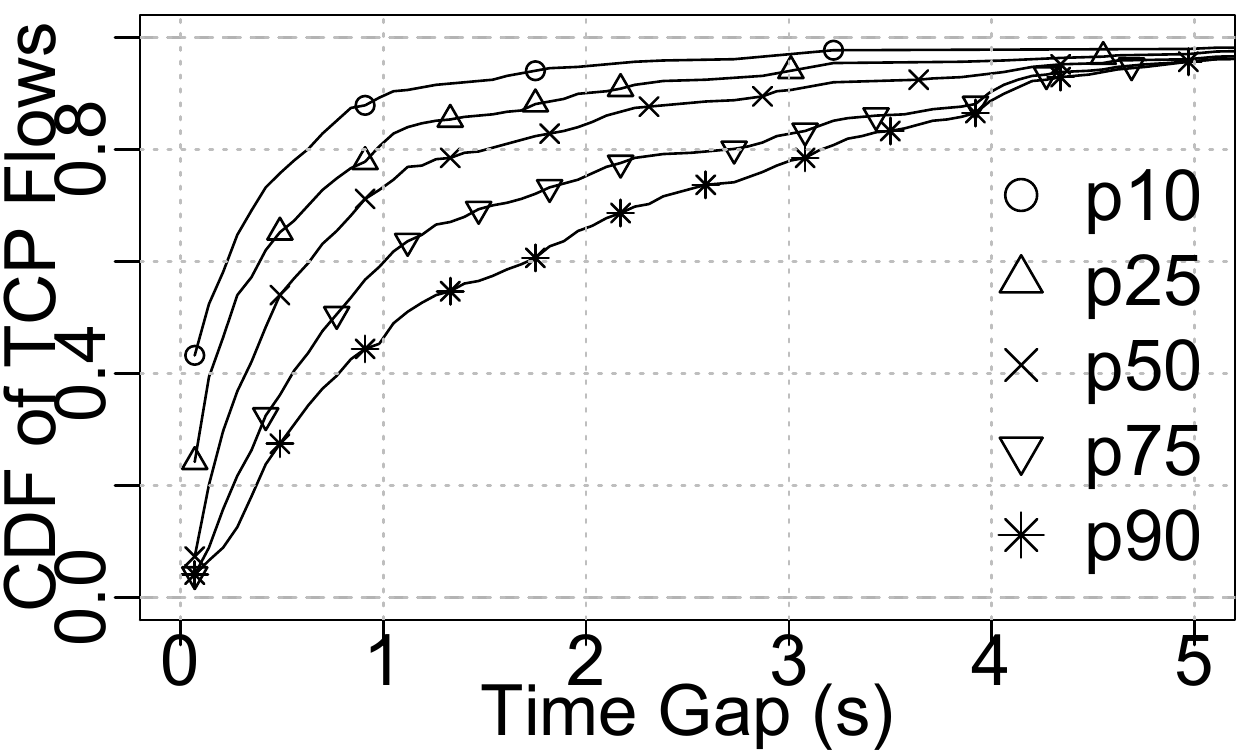}
\vspace{-18pt}
\caption{Distributions of time gaps in retransmissions.}
\label{fig:user_dist_gap}
\endminipage
\hfill
\minipage{0.24\textwidth}
  \includegraphics[width=\linewidth]{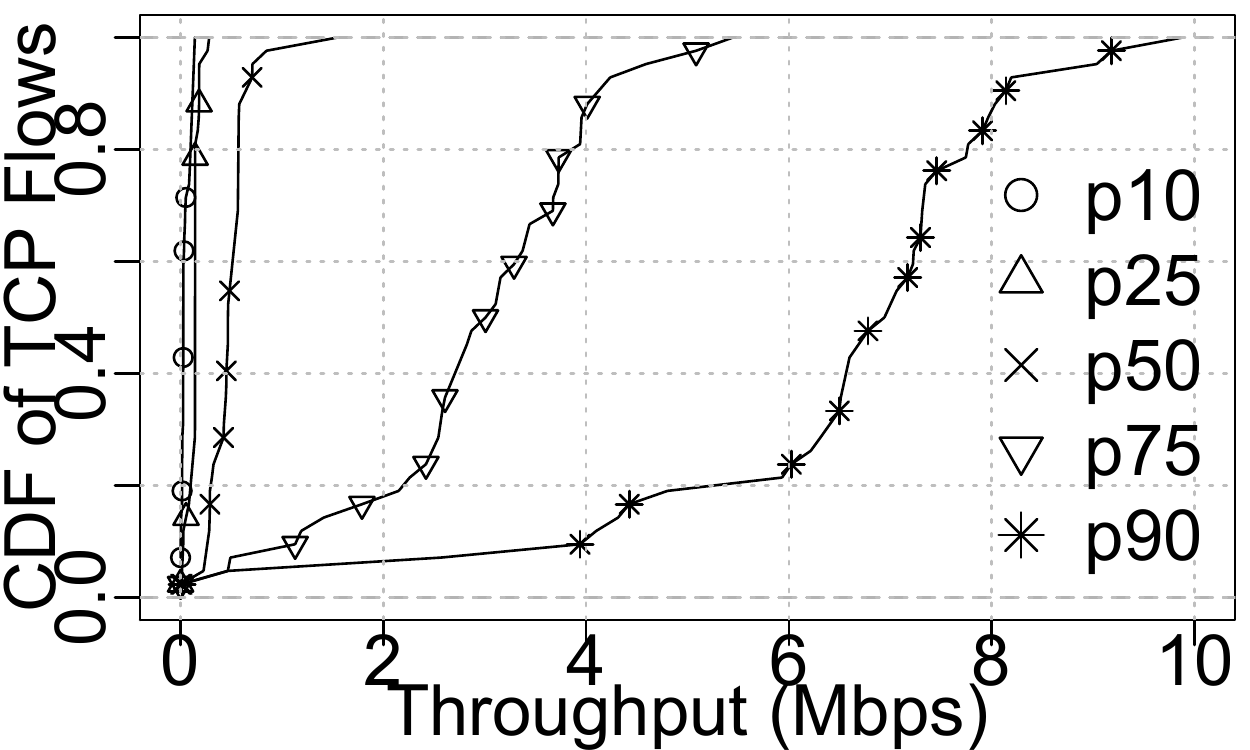}
\vspace{-18pt}
\caption{Distributions of estimated throughput.}
\label{fig:user_dist_thr}
\endminipage
\hfill
\minipage{0.24\textwidth}
  \includegraphics[width=\linewidth]{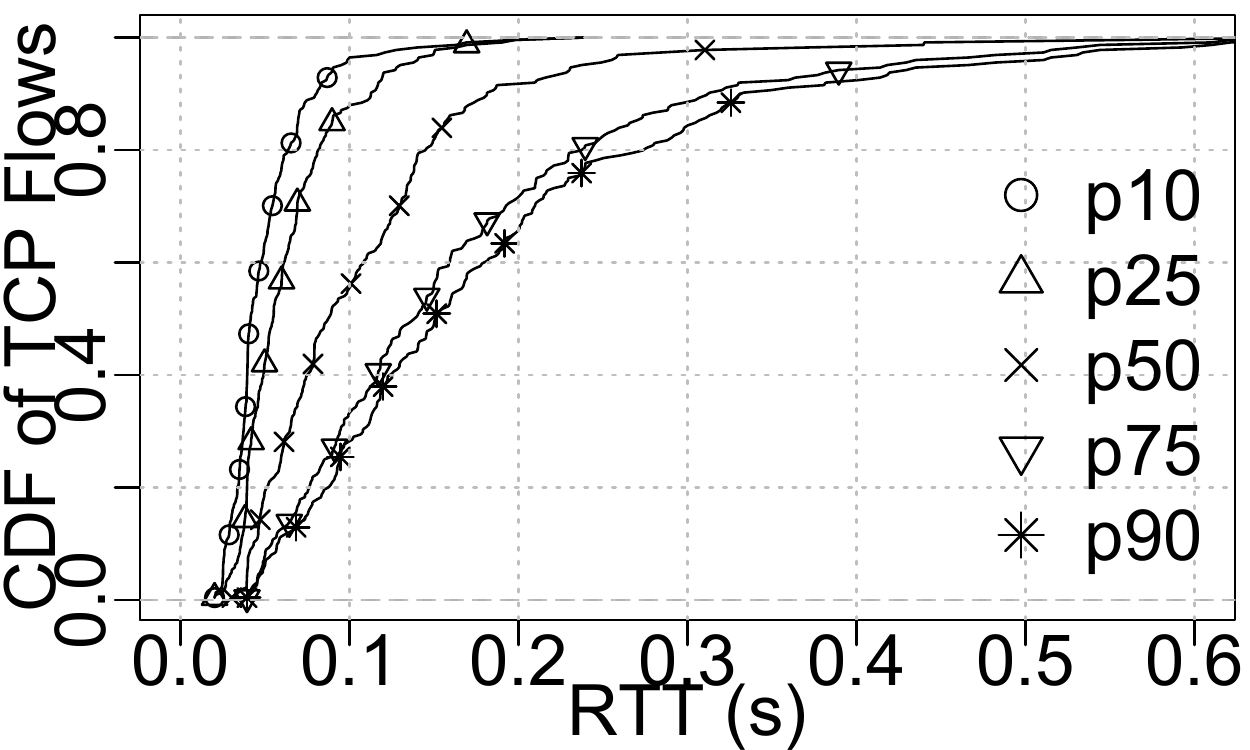}
\vspace{-18pt}
\caption{Distributions of RTT between ACK segments.}
\label{fig:user_dist_rtt}
\endminipage
\vspace{-10pt}
\end{figure*}

\vspace{-5pt}
\section{Cellular Network Characteristics}
\label{sec:setup}

The packet traces captured with \texttt{TCPDump} we use in our study  only indicate retransmissions by the server, not per se packet loss in the network. 
For example, while the server could retransmit due to socket timeouts caused by temporary congestion, or packet drop in the network due to corruption, TCP congestion control interprets both these events as congestion in the network and retransmits the segments it infers are lost~\cite{rfc5690}.
Our approach to identify loss relies on reverse engineering TCP's reaction to the changing dynamics of the cellular network. We argue that from the outside of a cellular network it is impossible to identify the real cause of a retransmission. 
From this point forward in the paper, we refer to packet loss as the TCP segments retransmitted by the server and use the terms loss and retransmission interchangeably.
\looseness -1

Next, we make four observations when analyzing the captured TCP traffic. 
The \textit{first observation} we make is that about 32\% of the total TCP connections in the chosen cellular network experience packet loss.
This observation is similar to our previous work on detecting TCP terminating proxies in cellular networks worldwide, where we use the packet loss information in the TCP logs provided by Linux kernel of production Akamai CDN servers~\cite{GoelMiddlebox16}.
\looseness -1

The \textit{second observation} we make is that losses in cellular networks are clustered.
In other words, when TCP interprets congestion, the server often retransmit many consecutive TCP segments.
In Figure~\ref{fig:cluster}, we show several boxplots, each for a 70\,ms slice, representing distributions of the number of TCP segments retransmitted by servers across all TCP connections.
Note that we only show number of retransmissions for those TCP connections that experienced retransmissions in the selected time slice on x-axis.
The x-axis represents the timestamp when a 70\,ms slice finishes.
The y-axis represents the number of packets retransmitted by server on a log scale.
From the figure we observe that many individual TCP connections experience clustered retransmissions.
For example, during the time slice finishing at 420\,ms, we observe that servers retransmit clusters of 5, 10, and even 20 segments for different TCP connections.
Although we observe clustered retransmissions for connections existing longer than 2.1\,seconds, for figure clarity we restrict the figure to the first 2.1\,seconds.
Our observations are similar to the recent work of Flach~\textit{et~al.}, where the authors observe that traffic policers introduce high packet loss rate in cellular networks~\cite{45411}.
\looseness -1

\vspace{-1pt}
The \textit{third observation} we make is that TCP connections experience retransmissions at multiple times during their respective lifetimes.
In Figure~\ref{fig:gap}, for each connection we show a boxplot distribution representing the time gaps between retransmission clusters observed in subsequent 70\,ms slices. 
Since we record the occurrence of retransmission clusters at every 70\,ms, each time gap is at least 70\,ms long.
From the figure we observe that for several connections the subsequent retransmission clusters appear within 500\,ms.
In other words, many connections experience clustered loss every half a second.
\looseness -1

\vspace{-1pt}
Finally, the \textit{fourth observation} we make is that when a TCP connection experiences clustered retransmissions, in a 70\,ms slice up to 40\% of the segments are retransmitted by the server.
We support this claim through Figure~\ref{fig:rtx}, where for each TCP connection we show a boxplot distribution of rate of retransmitted TCP segments, across different 70\,ms time slices.
Note that this retransmission rate is different from the aggregate retransmission that the connection experiences during its entire lifetime.
For example, while there could be 20\% segments retransmitted in a given 70\,ms time slice, however, the aggregate retransmission rate for the connection can be significantly lower than 20\%.
\looseness -1

\vspace{-8pt}
\section{Emulating Cellular Networks} 
\label{sec:emulation}
\vspace{-3pt}

One of the goals of this study is to improve existing techniques for emulating cellular networks.
While there exist many network emulators to model loss, latency, and bandwidth between clients and servers, such as TC~NETEM~\cite{tc}, Network Link Conditioner~\cite{linkConditioner}, these emulators introduce packet loss on network links at random times during emulation.  
Such emulators do not achieve a realistic emulation of cellular networks as loss in cellular networks does not occur at random times.
Using TC~NETEM and our observations from Section~\ref{sec:setup}, we develop a testbed that refines how packet loss is introduced on network links between clients and servers. 
\looseness -1

\begin{figure*}[t]
\centering
\minipage{0.325\textwidth}
  \includegraphics[width=\linewidth]{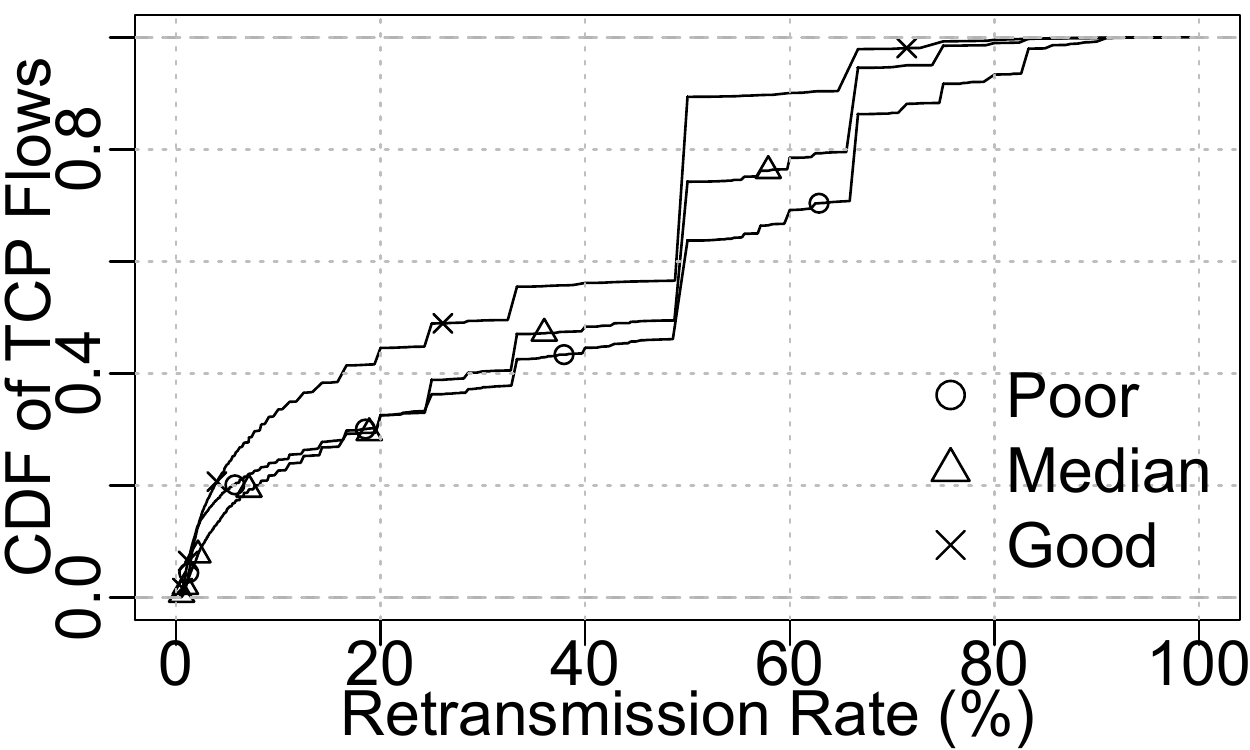}
\vspace{-15pt}
\caption{Distributions of retransmission rates.}
\label{fig:tcp_dist_rtx}
\endminipage
\hfill
\minipage{0.325\textwidth}
  \includegraphics[width=\linewidth]{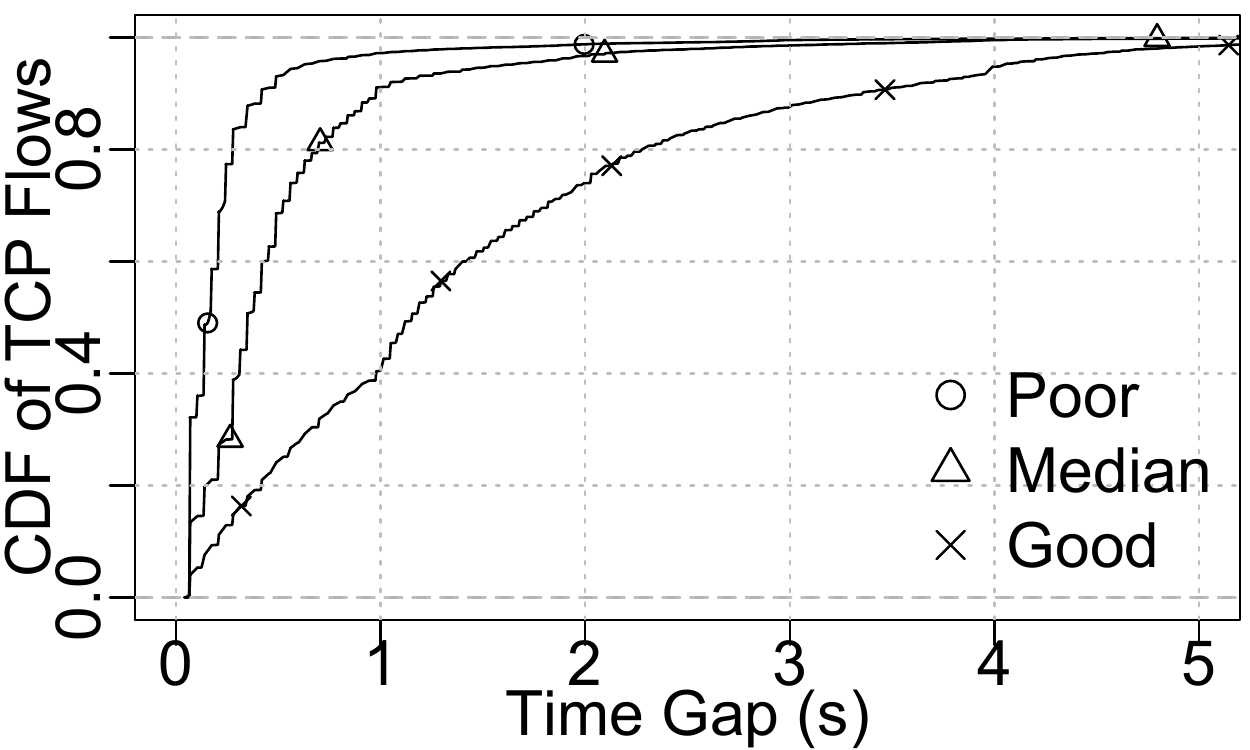}
\vspace{-15pt}
\caption{Distributions of time gap between retransmission clusters.}
\label{fig:tcp_dist_gap}
\endminipage
\hfill
\minipage{0.325\textwidth}
  \includegraphics[width=\linewidth]{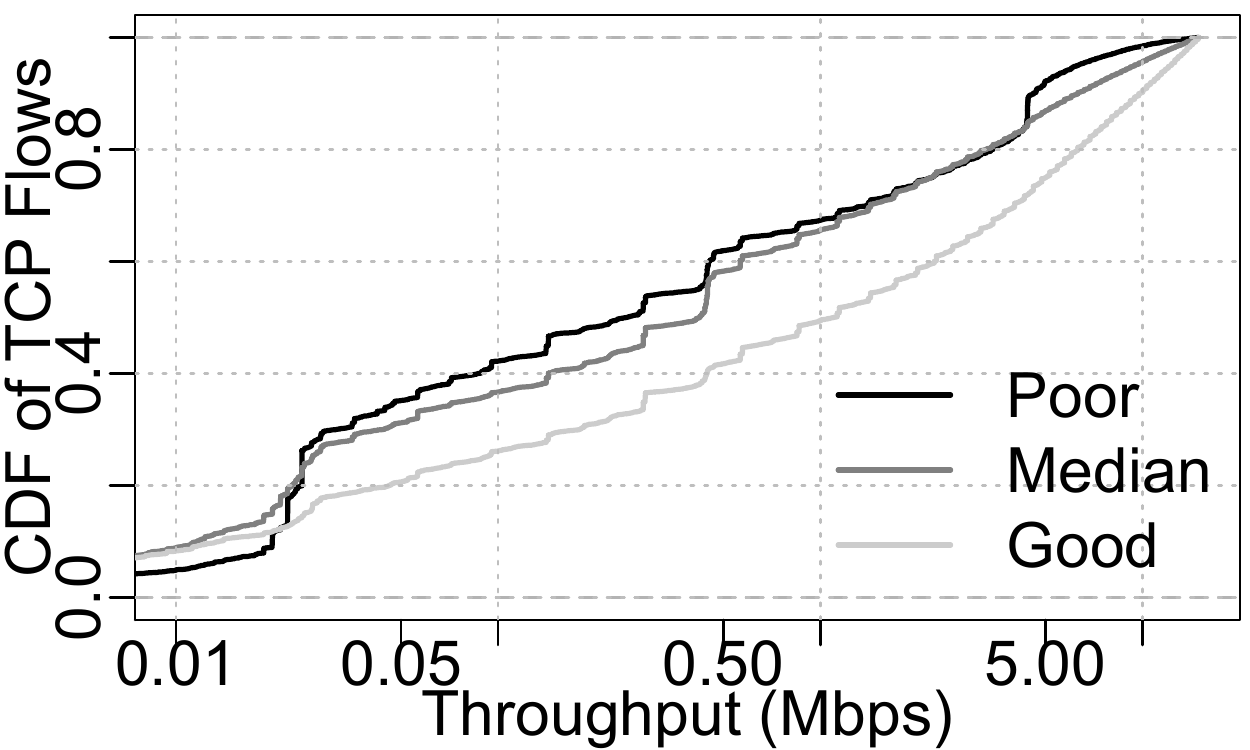}
\vspace{-15pt}
\caption{Distributions of estimated throughput.}
\label{fig:tcp_dist_thr}
\endminipage
\vspace{-15pt}
\end{figure*}

Next, using only the captured TCP connections that experienced loss~(32\% of the collected dataset), we develop two emulation scenarios: 1)~\emph{network quality} observed in different situations, such as good or bad reception, stationary or moving user; and 2)~\emph{time gap between retransmission clusters}, or how frequently a TCP connection experiences loss.
For the first emulation scenario, we classify all TCP connections into four different network qualities, namely \textit{Good}, \textit{Fair}, \textit{Passable}, and \textit{Poor}.
In the second emulation scenario, we classify all TCP connections into three different categories based on the \emph{time gap between retransmission clusters}, or how frequently a TCP connection experiences loss, namely \textit{Good}, \textit{Median}, and \textit{Poor}.
\looseness -1

\vspace{-8pt}
\subsection{Classifying Connections Based on Network Quality}
\label{sec:quality}
In Figures~\ref{fig:user_dist_rtx}-\ref{fig:user_dist_rtt}, we show distributions of 10\textsuperscript{th}, 25\textsuperscript{th}, 50\textsuperscript{th}, 75\textsuperscript{th}, and 90\textsuperscript{th} percentile retransmission rate, time gap between retransmission clusters, estimated throughput, and RTT, as observed across all TCP connections in all 70\,ms time slices.
Next, using these distributions, we label TCP connections into the five different categories, as shown in Table~\ref{tbl:emulation}. 
For example, when emulating a network with \textit{Good} quality, we select the 10\textsuperscript{th} percentile~(p10) retransmission rate distribution from Figure~\ref{fig:user_dist_rtx}, 10\textsuperscript{th} percentile RTT~(p10) distribution from Figure~\ref{fig:user_dist_rtt}, 90\textsuperscript{th} percentile~(p90) throughput distribution from Figure~\ref{fig:user_dist_thr}, and 90\textsuperscript{th} percentile~(p90) distribution of time gap between retransmission clusters from Figure~\ref{fig:user_dist_gap}.
We then model the network between the client and server by modifying the above characteristics every 70\,ms.
\looseness -1

\vspace{6pt}
\noindent
\textit{Note:} Network bandwidth in cellular networks is attributed to the base station and is
not dependent on loss and RTT; however, network throughput is dependent on loss and RTT~\cite{sprout}.
In our emulation, for lack of actual bandwidth information, we use the observed throughput to model the bandwidth. 
The collected TCP traces do not reveal the network bandwidth and therefore one can only calculate the achieved throughput.
We acknowledge the fact that using observed throughput as a substitute for bandwidth is not ideal. 
On the other hand, modeling the bandwidth with a constant value for the entire duration of the emulation is also not realistic.
Therefore, in order to model the bandwidth, we decide to use the observed throughput as an approximate value for network bandwidth. 
\looseness -1

\begin{table}[t]
\centering
\resizebox{0.945\columnwidth}{!}{
\begin{tabular}{l|cccc}
\multicolumn{1}{c|}{\textbf{Quality}} & \textbf{Rtx. Rate} & \textbf{Time Gap} & \textbf{Throughput} & \textbf{RTT} \\
\hline\hline
\textbf{Good}   & p10           & p90               & p90                 & p10          \\
\textbf{Fair}        & p25           & p75               & p75                 & p25          \\
\textbf{Passable}      & p50           & p50               & p50                 & p50          \\
\textbf{Poor}        & p75           & p25               & p25                 & p75          \\
\hline
\end{tabular}
}
\vspace{-2pt}
\caption{Emulation based on network quality.}
\label{tbl:emulation}
\vspace{-16pt}
\end{table}

\vspace{-7pt}
\subsection{Classifying Connections based on Time Gap Between\\ Retransmission Clusters} 
\label{sec:condition}

When using this classification to categorize TCP connections, a \textit{Poor} condition represents a network where subsequent retransmission clusters are separated by less than 250\,ms in the median case.
Next, a \textit{Median} condition represents a network where subsequent retransmission clusters occur every 250\,ms to 750\,ms in the median case.
And finally, a \textit{Good} condition represents a network where subsequent retransmission clusters are separated by atleast 750\,ms in the median case.
\looseness -1

\begin{figure}[t]
\centering
\minipage{0.235\textwidth}
  \includegraphics[width=\linewidth]{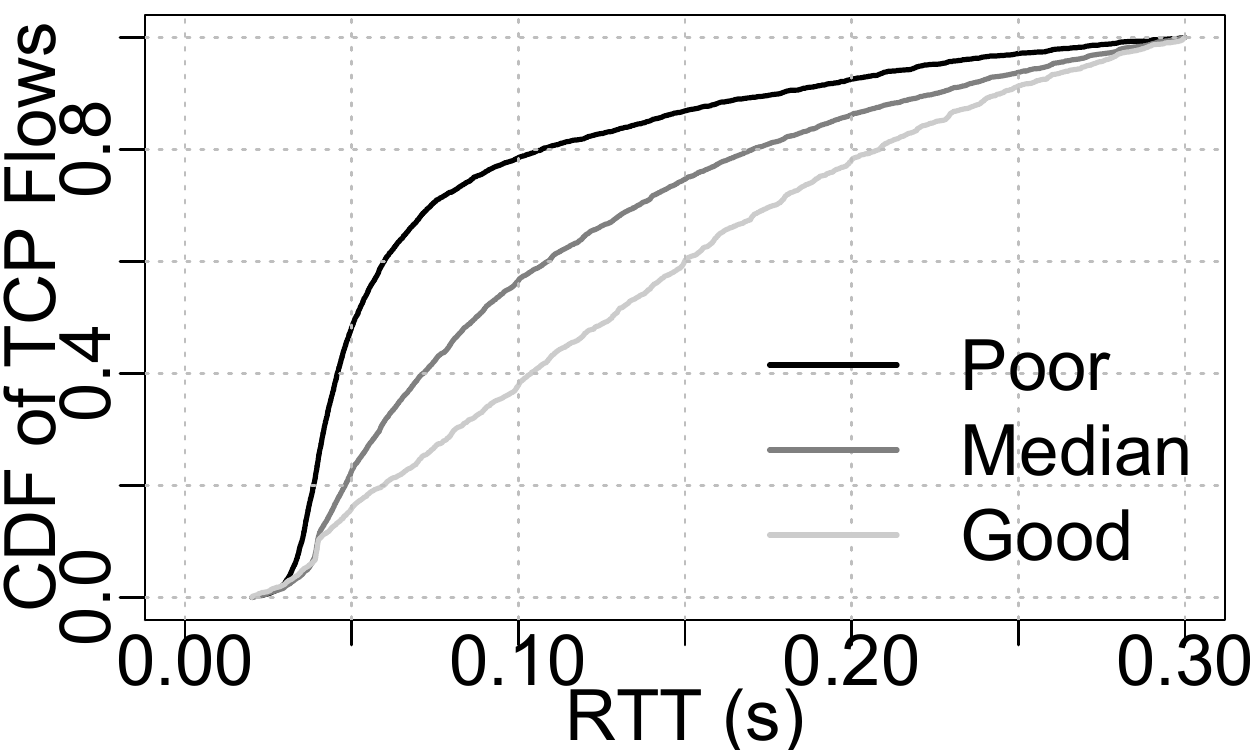}
\vspace{-15pt}
\caption{RTT Distributions according to \texttt{tshark}.}
\label{fig:tcp_dist_rtt}
\endminipage
\hfill
\minipage{0.235\textwidth}
  \includegraphics[width=\linewidth]{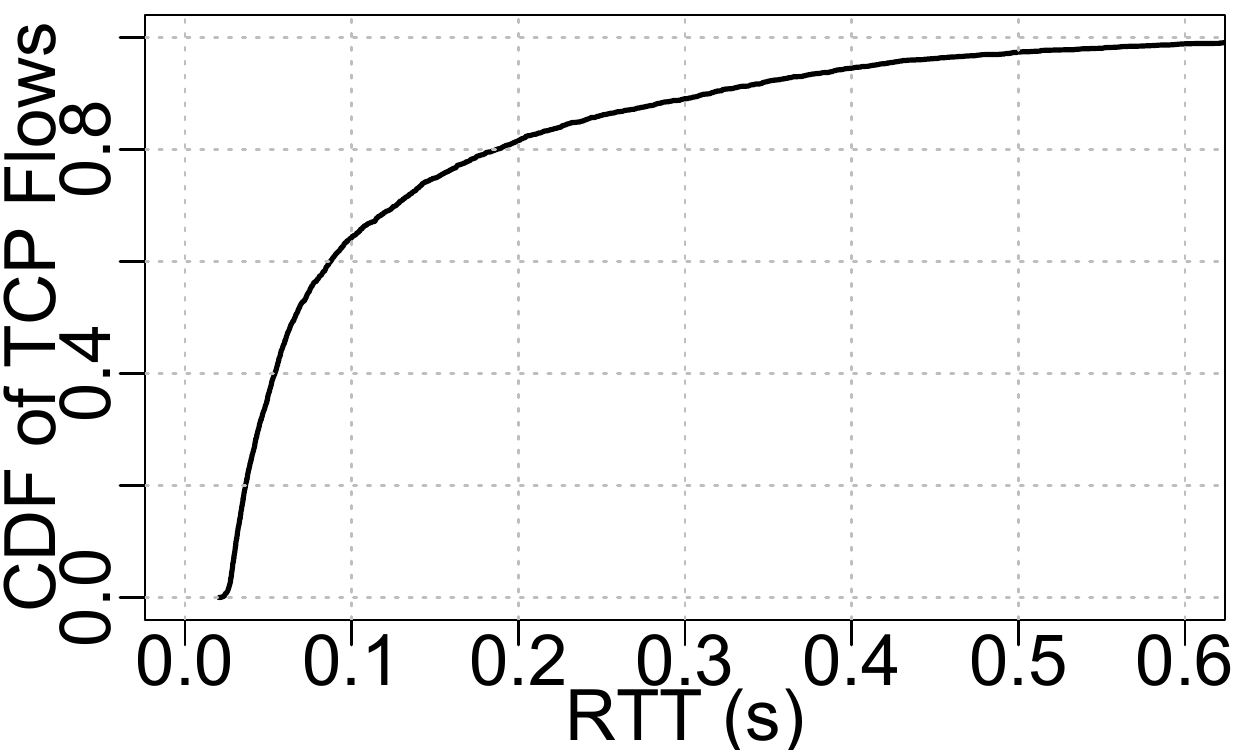}
\vspace{-15pt}
\caption{RTT Distributions from the kernel TCP logs.}
\label{fig:tcp_dist_global_rtt}
\endminipage
\vspace{-18pt}
\end{figure}

In Figures~\ref{fig:tcp_dist_rtx}-\ref{fig:tcp_dist_thr}, we show distributions of retransmission rate, time gap between retransmission clusters, and estimated throughput, for different classifications of cellular network conditions.
For example, in Figure~\ref{fig:tcp_dist_gap} we show that TCP connections that belong to a \textit{Good}, \textit{Median}, and \textit{Poor} network conditions experience retransmission clusters after 1.15 seconds, 350\,ms, and 165\,ms respectively in the median case.
\looseness -1

With respect to RTT distributions, from Figure~\ref{fig:tcp_dist_rtt} we observe that RTT for \textit{Poor} networks is much lower than RTT for \textit{Good} conditions.
We argue that this behavior is due to the fact that when client receives out-of-order segments, TCP congestion control on the client transmits \texttt{TCP ACKs} immediately to accelerate recovery of lost segments~\cite{rfc2581}.
Therefore, a receiver on a \textit{Poor} network (experiencing loss frequently) sends \texttt{ACK}s as soon as it receives an out-of order packet. 
When such \texttt{ACK}s are received by the server, \texttt{tshark} calculates the time difference between the previous and the new \texttt{ACK} segments, which results in exactly one RTT as \texttt{ACK}s are sent immediately following a loss. 
In other words, if segment X is lost and server keeps sending X+N segments, the client will send a \texttt{DUP ACK} immediately for every N packets until X is recovered.
\looseness -1
 
 \begin{figure*}[t]
\centering
\minipage{0.28\textwidth}
  \includegraphics[width=\linewidth]{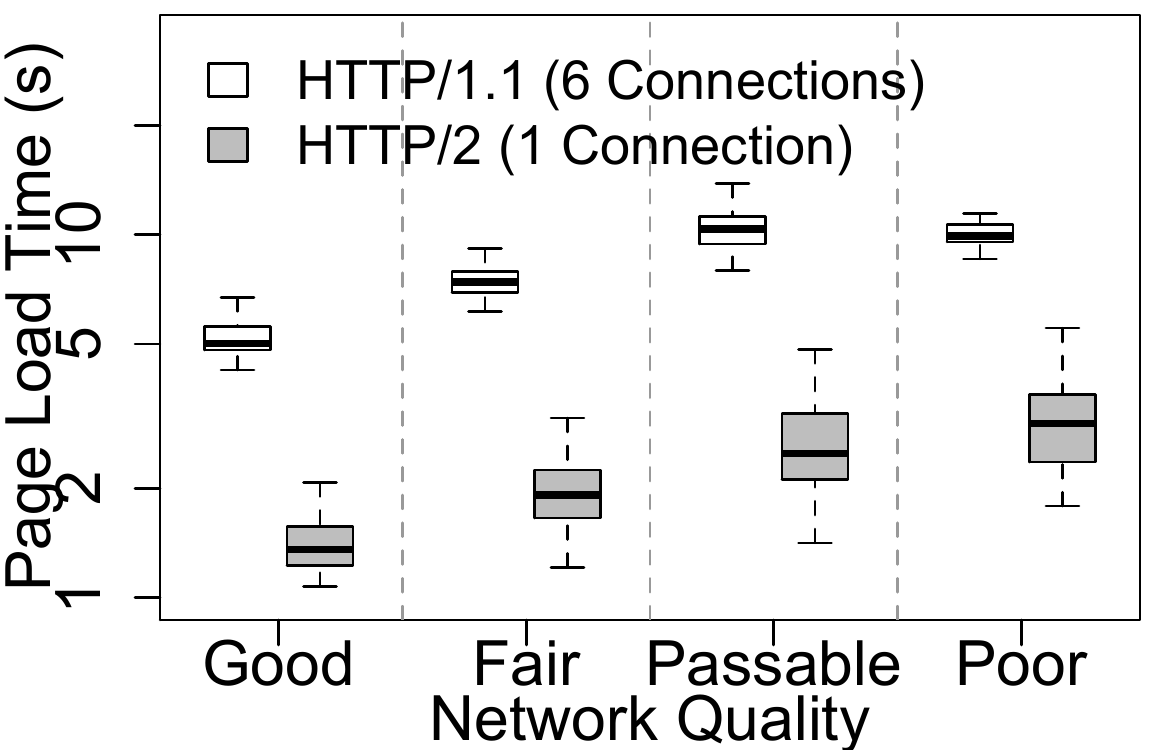}
\vspace{-15pt}
\captionsetup{width=0.9\columnwidth}
\caption{PLTs of a webpage with 365 objects of size 1\,KB.}
\label{fig:spinning}
\endminipage
\hfill
\minipage{0.28\textwidth}
  \includegraphics[width=\linewidth]{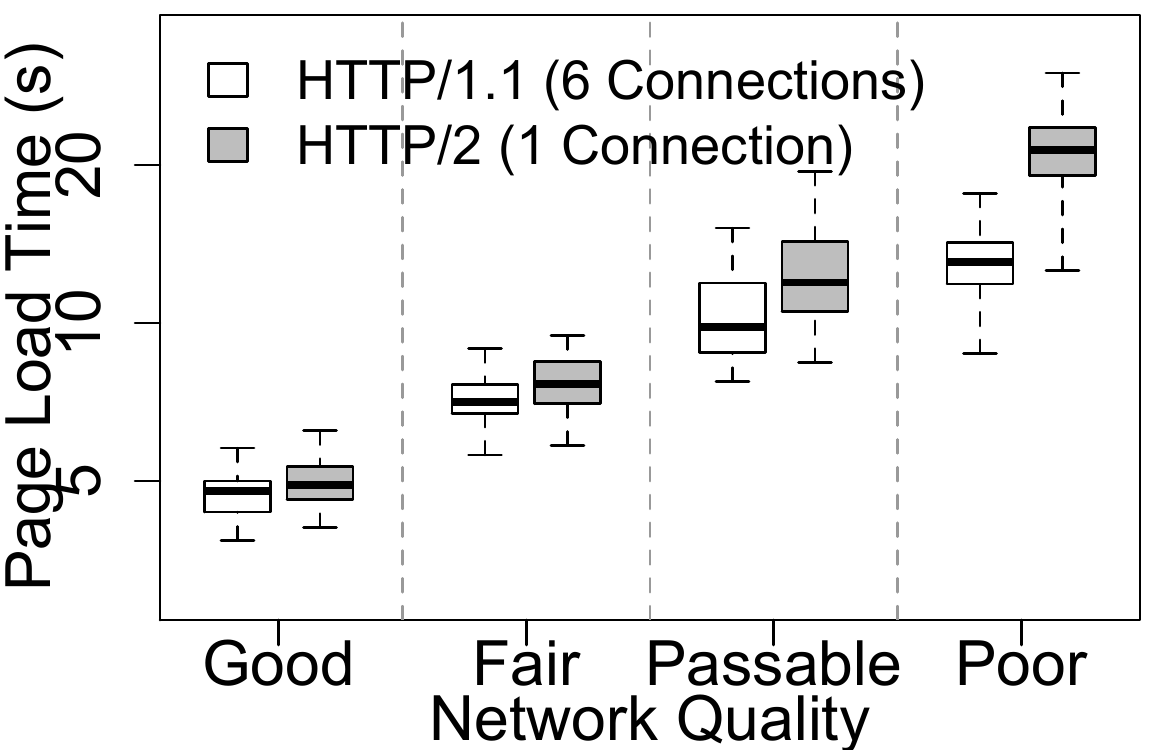}
\vspace{-15pt}
\captionsetup{width=0.9\columnwidth}
\caption{PLTs of a webpage with 10 objects of size 435\,KB each.}
\label{fig:moritz}
\endminipage
\hfill
\minipage{0.425\textwidth}
  \includegraphics[width=\linewidth]{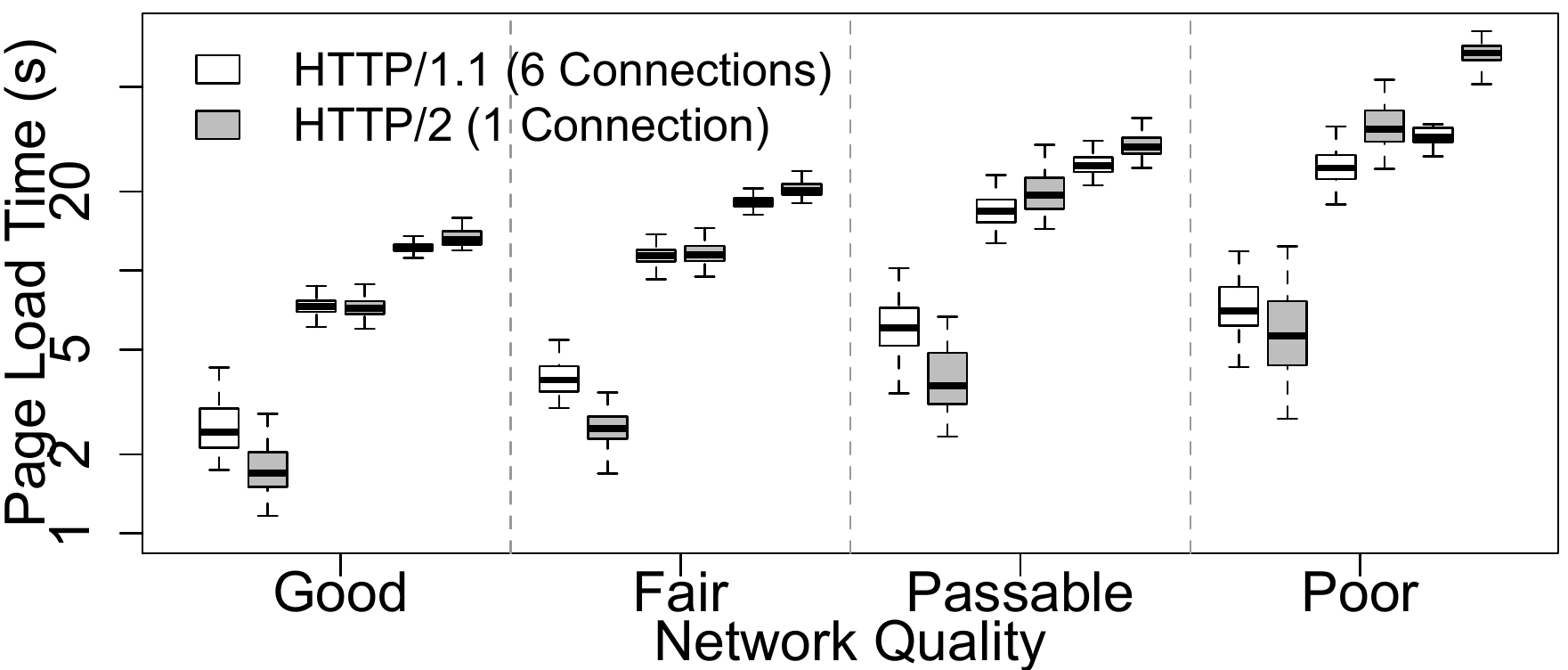}
\vspace{-15pt}
\captionsetup{width=0.9\columnwidth}
\caption{PLTs of webpages with 136 objects of size 1-620\,KB each.}
\label{fig:javier}
\endminipage
\vspace{-15pt}
\end{figure*}

In the case of \textit{Good} connections~(experiencing loss rarely), the TCP congestion control on the client sends \texttt{ACK}s for only every other TCP segment.
Moreover, the client waits up to 500\,ms to accumulate all \texttt{ACK}s and then transmit an \texttt{ACK} for the most recently in-order segment received~\cite{rfc2581}.
Therefore, when there is no loss for several RTTs, the time difference between two received \texttt{ACK} segments is much larger than the true RTT between the client and server. 
In other words, if the client delays \texttt{ACK}s, the \texttt{ACK} RTT value calculated by \texttt{tshark} will result in higher values than the values calculated in the case of a \textit{Poor} network.
Note that even though recent versions of Red Hat only delay \texttt{ACK}s for 40\,ms, our argument, related to why calculating RTTs based on the arrival times of \texttt{ACK}s is inaccurate, still holds valid.
Specifically, if the TCP stack on the client delays the transmission of \texttt{ACK}s by 40\,ms, the overall RTT observed by the server is more than 150\%~(110\,ms) of the true RTT~(70\,ms in the median case).
\looseness -1

Therefore, to extract accurate RTT distributions for different network conditions, we extract over 15\,million TCP log lines recorded by the same CDN cluster over a week, where each TCP log line represents an end-to-end connection between a client and server's port 443. 
We are interested to understand whether or not latency is correlated to loss in cellular networks.
From each TCP log line we extract the minimum, average, maximum RTT, and the time to establish TCP connection, along with the total loss the connection experienced during its lifetime.
We observe that latency in cellular networks has an extremely low correlation with the observed loss.
Specifically, for 15\,million TCP connections, the correlation values between the overall observed loss and minimum, maximum, average RTT, and the time to establish TCP connection were only 0.004, 0.17, 0.24, and 0.05 respectively.
This is likely due to the fact that latency in cellular networks depends on multiple factors such as, time spent in radio resource negotiations, time to switch the device radio from idle to active, packet queuing on routers in the core network.
Therefore, given the low correlation between latency and loss, from the TCP logs we select the time to establish TCP connections as a standalone distribution of RTT to use when emulating \textit{Good}, \textit{Median}, and \textit{Poor} network conditions.
We show the RTT distribution in Figure~\ref{fig:tcp_dist_global_rtt}, which has the median RTT of 70\,ms, same as the one we used for slicing TCP connections.
\looseness -1

\vspace{-12pt}
\subsection{Emulation in action}

To emulate networks based on the two techniques described in Sections~\ref{sec:quality} and~\ref{sec:condition}, we setup a network topology using three machines with TCP~CUBIC installed on Ubuntu 14.04.
On the first machine, we configure a \texttt{client} that runs Chromium Telemetry for loading webpages using the Google Chrome browser~\cite{telemetry}.
On the second machine, we configure an Apache Web \texttt{server} that supports \texttt{h1} and \texttt{h2} on different virtual hosts.
Finally, on the third machine, we configure a \texttt{bridge} to connect the client and the server. 
For our experiments, the initial congestion window~(ICW) on the server is set to 10 segments of size 1460 bytes each.
The \textit{receive windows} advertised by the \texttt{client} and \texttt{server} during connection setup are set to 65\,KB.
\looseness -1

Depending on the emulation scenarios defined in Sections~\ref{sec:quality} and~\ref{sec:condition}, we configure the \texttt{bridge} to use TC~NETEM commands to modify loss, time to subsequent loss, RTT, and bandwidth every 70\,ms.
Note that we use the retransmission rates discussed in Sections~\ref{sec:quality} and Section~\ref{sec:condition} to model loss rates on the \texttt{bridge}.
Similarly, we use the time gap between retransmission clusters to model the time when loss is introduced on the link between the client and server.
Finally, we use the estimated throughput to model network bandwidth between the client and server, with the reasoning described in notes of Section~\ref{sec:quality}.
\looseness -1

\vspace{-5pt}
\section{Comparing Web performance of \texttt{h2} and \texttt{h1}}
\label{sec:results}

In this section, we emulate the five network qualities described in Section~\ref{sec:quality} to compare PLTs over \texttt{h2} and \texttt{h1}.
Since there is no standard definition of PLT, similarly to many other studies~\cite{Erman:2013:TSM:2535372.2535399,GoelIPv6,7288411,7557456,DBLP:journals/corr/SteinerG16,pamhttp2,Wang:2014:SS:2616448.2616484}, we estimate the PLT as the time from when the user enters the URL in the Web browser until the browser fires the JavaScript's \textit{OnLoad} event.
For measuring the PLT, we use the \texttt{client} to load several webpages synthesized from HTTP Archive~\cite{httparchive}.
HTTP Archive is a repository that maintains structures of many popular webpages designed for both mobile and desktop screens.
\looseness -1

The webpages we synthesize represent many popular mobile websites, ranging in the HTML document size, number of embedded objects, and total webpage size.
Note that we use synthesized webpages instead of relying on real webpages, because real webpages contain many third party objects that could influence the overall PLT by up to 50\%~\cite{GoelThird}.
Further, many third party objects are downloaded over \texttt{h1}, even though the base page HTML could be downloaded over \texttt{h2}.
As such, third party objects on real webpages introduce interference in PLT estimations.
Therefore, we synthesize webpages that do not include third party content and whose structure, object size, and overall size represent popular webpages.
\looseness -1

In our experiments, for each scenario, we load each page 200 times over \texttt{h2} and 200 times over \texttt{h1} using TLS.
Note that we use log y-axis on all figures for clarity.
\looseness -1

The \textit{first webpage} contains 365 objects of size 1\,KB each, the HTML document size of about 38\,KB, and a total page size of about 400\,KB.
This webpage represents 40\% of the top 1000 mobile webpages that embed up to 400 objects, 25\% of the webpages with up to 40\,KB HTML document size, 52\% of the webpages with a total webpage size of up to 1\,MB, and 49\% of the pages that transfer up to 400\,KB of image data~\cite{httparchive}.
In Figure~\ref{fig:spinning}, we observe that PLTs over \texttt{h2} are significantly lower than PLTs over \texttt{h1}.
This is because \texttt{h1} establishes six TCP connections with the server, which allows the server to transfer only six objects in parallel.
As each TCP connection for \texttt{h1} can only download one object at a time, the server sends a total of 6 segments~(6\,KB) for the requested six objects, before it waits for the next request. 
In other words, \texttt{h1} can only send 6\,KB of data in each round trip, regardless of the congestion window size on the server. 
On the other hand, unlike \texttt{h1}, \texttt{h2} multiplexes many objects on the single TCP connection and can pack multiple objects in one TCP segment.
Initially, a single TCP connection for \texttt{h2} allows the server to send 10 segments, that is 14.6\,KB~(14 objects for this site), with the number of segments that the server can send growing exponentially with each round trip during the TCP slow start.
Therefore, a webpage with many small objects make an ideal case for \texttt{h2} to reduce the number of round trips required, in comparison to \texttt{h1}.
In fact, since \texttt{h1} requires more round trips to load this webpage, it experiences more aggregate packet loss compared to \texttt{h2}. 
Finally, when packet loss forces the server to drop its TCP congestion window, the server in the case of \texttt{h2} still transmits more objects than in the case of \texttt{h1}.
Therefore, we observe that \texttt{h2} outperforms \texttt{h1} across all emulated scenarios. 
\looseness -1

The \textit{second webpage} contains 10 large objects of size 435\,KB each, the HTML document size of about 10\,KB, and total page size of about 4\,MB.
This webpage represents 40\% of the top 1000 mobile webpages that embed up to 400 objects, 38\% of the webpages with up to 20\,KB HTML document size, and 6\% of the webpages with a total webpage size of about 4\,MB~\cite{httparchive}.
From Figure~\ref{fig:moritz}, we observe that \texttt{h1} outperforms \texttt{h2} across all emulated network qualities.
We argue that since the object sizes in this page are much larger than the object sizes in the previous page, the server in the case of \texttt{h1} uses all six TCP connections to send a total of about 60 segments~(87.6\,KB) in the first round trip.
Whereas, in the case of \texttt{h2} with one TCP connection, the server sends only 10 segments~(14.6\,KB) of data in the first round trip. 
Note that the number of segments that the server can send over each connection doubles in every round trip.
During TCP slow start, the cumulative congestion window usable over \texttt{h1} is six times larger than the congestion window usable over \texttt{h2}.
Further, as the network quality gets worse when loss occurs, the congestion window of the single connection over \texttt{h2} does not grow as much as it grows cumulatively for six connections in the case of \texttt{h1}.
For example, in \textit{Poor} network quality we observe that PLTs over \texttt{h2} are significantly higher than PLTs over \texttt{h1}.
Therefore, a webpage with many large objects requires many more round trips over \texttt{h2} than it would require over \texttt{h1}.
\looseness -1

The HTTP Archive data also suggests that many popular webpages embed both small and large objects, however, their counts differ significantly.
Therefore, using the HTTP Archive data we synthesize three more webpages, each containing 136 objects of size in the range of 1\,KB to 620\,KB, but with different number of large objects.
Specifically, the first webpage is of size 2\,MB, with three large objects~(ranging from 30\,KB to 620\,KB in size) and 133 small objects~(ranging from 20\,B to 5\,KB in size).
This webpage is similar to the 16\% of the top 1000 mobile webpages depicting many popular e-commerce, news, and sports websites~\cite{httparchiveDownloads}.
The second webpage is of size 8\,MB, with 12 large objects and 124 small objects.
This webpage is similar to many news, airlines, and blogging websites, such as WikiHow, NYPost, Delta, and others~\cite{httparchiveDownloads}.
The third webpage is of size 12\,MB, with 18 large objects and 118 small objects.
This webpage is similar to many news, online gaming, and streaming websites, such as Twitch.tv, TomsHardware.com, and others~\cite{httparchiveDownloads}.
The first, second, and third pairs of boxplots in Figure~\ref{fig:javier} for each network quality represent the distribution of PLTs for webpages of size 2\,MB, 8\,MB, and 12\,MB, respectively.
From the figure we observe that for a 2\,MB page, PLTs over \texttt{h2} are lower than PLTs over \texttt{h1}.
This is because, similarly to Figure~\ref{fig:spinning}, server sends many more objects over \texttt{h2} in parallel, compared to only six objects that the server sends in parallel over \texttt{h1}.
\looseness -1

\begin{figure}[t]
\centering
\minipage{0.35\textwidth}
  \includegraphics[width=\linewidth]{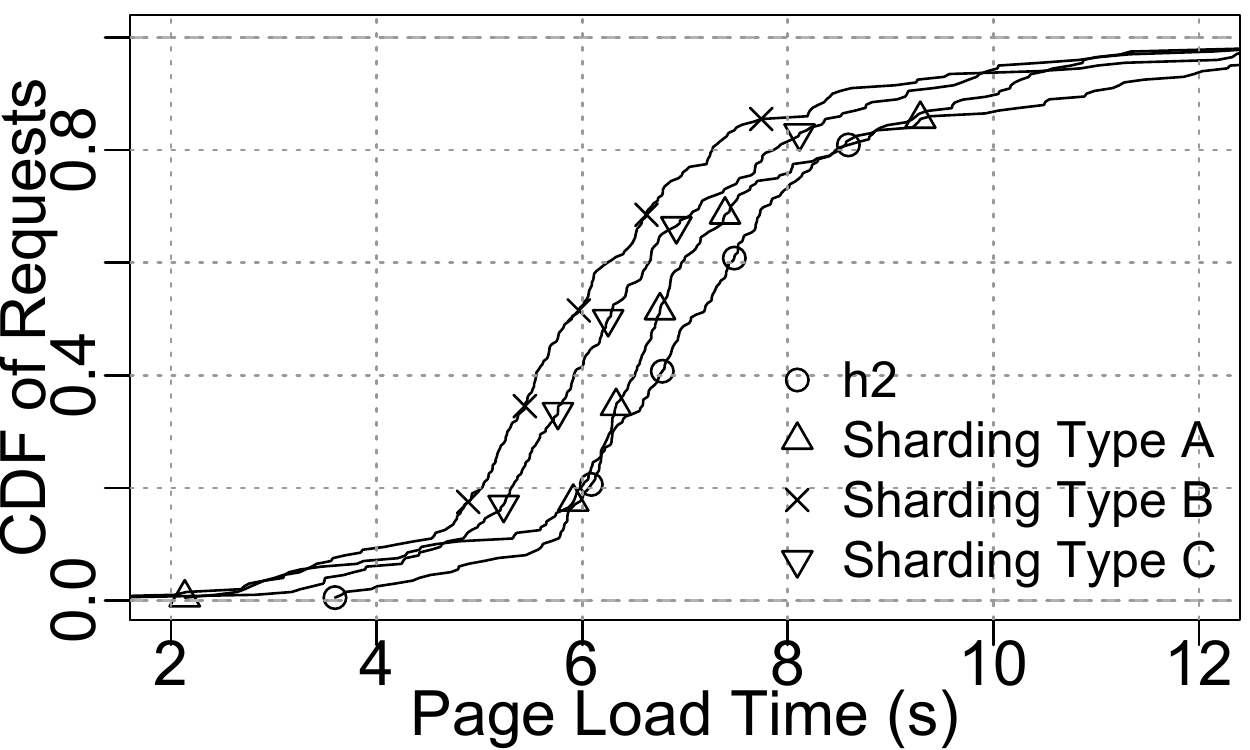}
\vspace{-15pt}
\captionsetup{width=1.2\columnwidth}
\caption{\texttt{h2} PLTs with different connection counts.}
\label{fig:sharding_effect}
\endminipage
\vspace{-15pt}
\end{figure}

For the 8\,MB page, although the page contains 12 large objects, PLTs over \texttt{h2} and \texttt{h1} are comparable under \textit{Good} and \textit{Fair} networks.
This is because \texttt{h2} gets benefits of multiplexing small-sized objects, whereas, \texttt{h1} suffers from HOL blocking for both small and large-sized objects.
However, as the network quality get worse, PLTs over \texttt{h2} become larger than \texttt{h1}, because in the case of \texttt{h2}, downloading large objects over single TCP connection suffers from small congestion window due to frequent packet loss.
In fact, a large object download multiplexed with small objects prevents the small objects from being downloaded in fewer round trips, especially when packet loss occurs during large object downloads.
\looseness -1

Finally, for the 12\,MB page, the PLTs over \texttt{h2} are always higher than \texttt{h1}.
This is because, the majority of the time is spent loading the large objects.
Therefore, when loss occurs, the congestion window on the server in the case of \texttt{h2} does not grow as much and as fast as it grows cumulatively for six connections in the case of \texttt{h1} -- affecting the PLTs over \texttt{h2} when downloading large objects.
Note that TCP's slow start is less important here as most of the PLT comes from the congestion avoidance phase.
To confirm whether the ICW impacts PLTs for large webpages, we loaded a webpage 25\,MB in size~(results not shown in the paper) and observed no statistical significance in the difference between \texttt{h2} and \texttt{h1} PLTs. 
\looseness -1

\begin{figure*}[t]
\centering
\minipage{0.325\textwidth}
  \includegraphics[width=\linewidth]{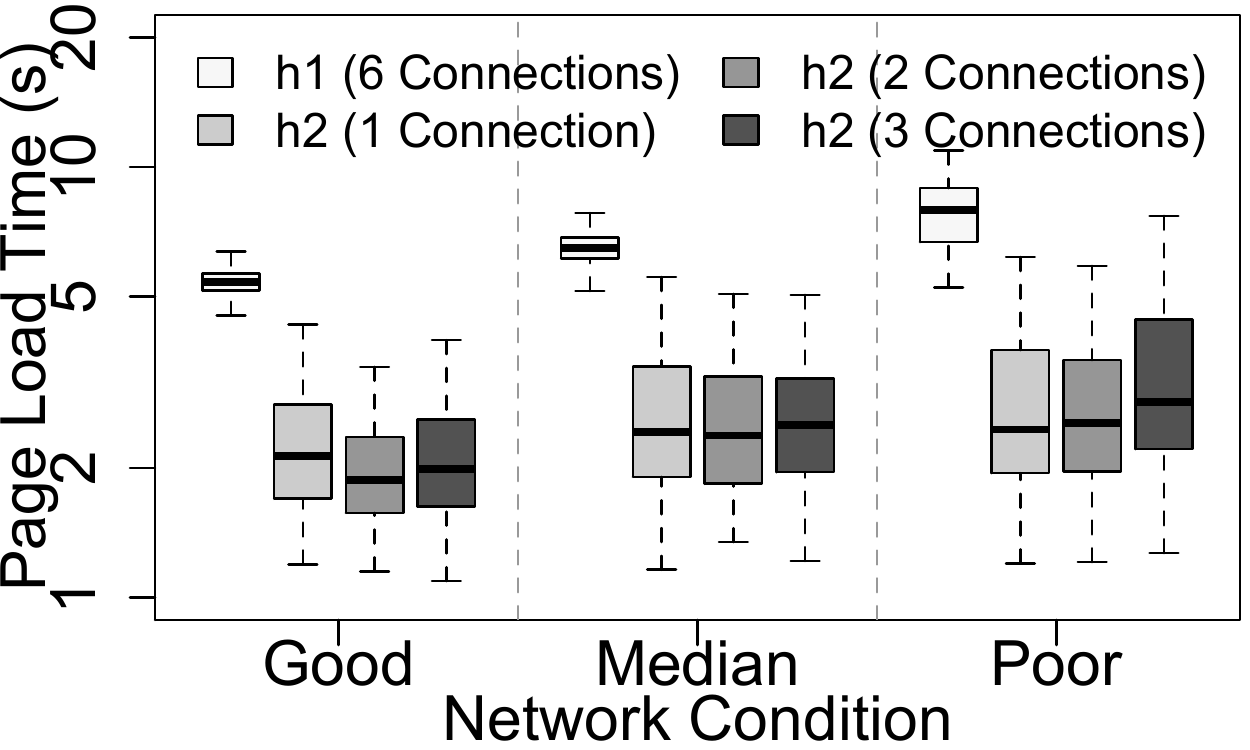}
\vspace{-15pt}
\caption{PLT distributions of a webpage with 365 objects of size 1\,KB each when emulating based on TCP loss frequency.}
\label{fig:spinning_shard}
\endminipage
\hfill
\minipage{0.325\textwidth}
  \includegraphics[width=\linewidth]{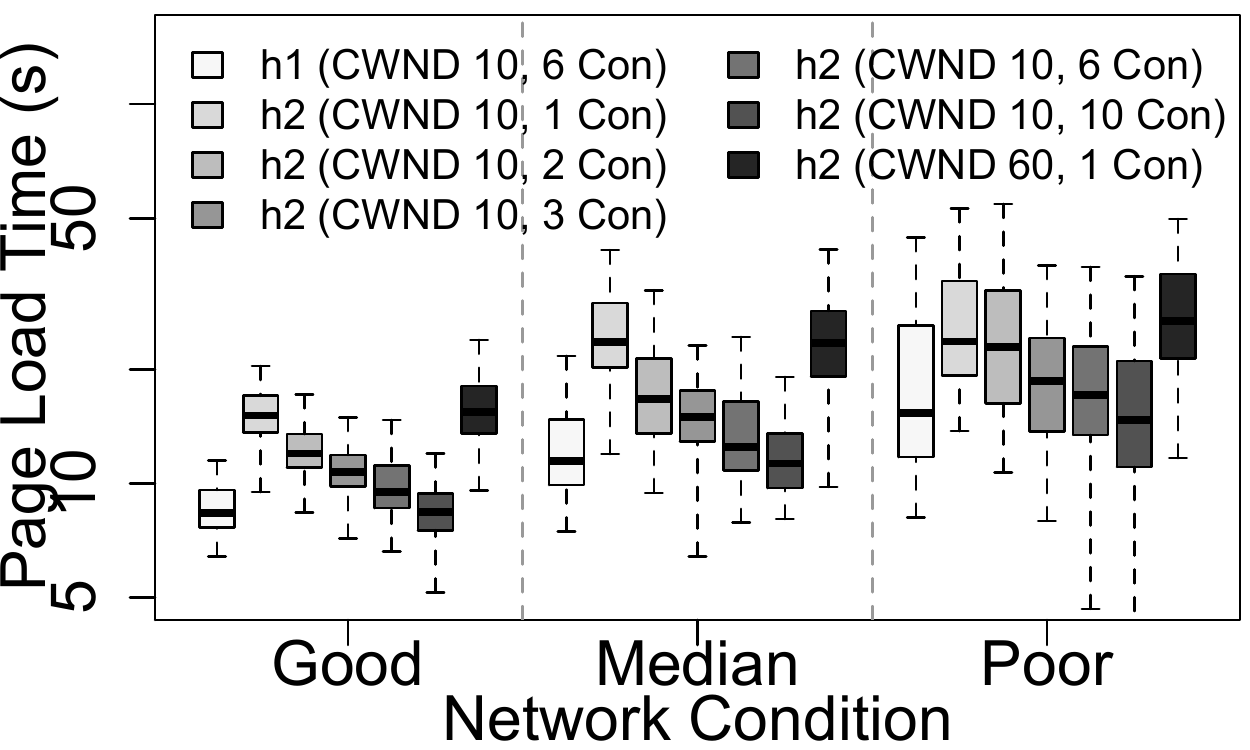}
\vspace{-15pt}
\caption{PLT distributions of a webpage with 10 objects of size 435\,KB each when emulating based on TCP loss frequency.}
\label{fig:moritz_shard}
\endminipage
\hfill
\minipage{0.325\textwidth}
  \includegraphics[width=\linewidth]{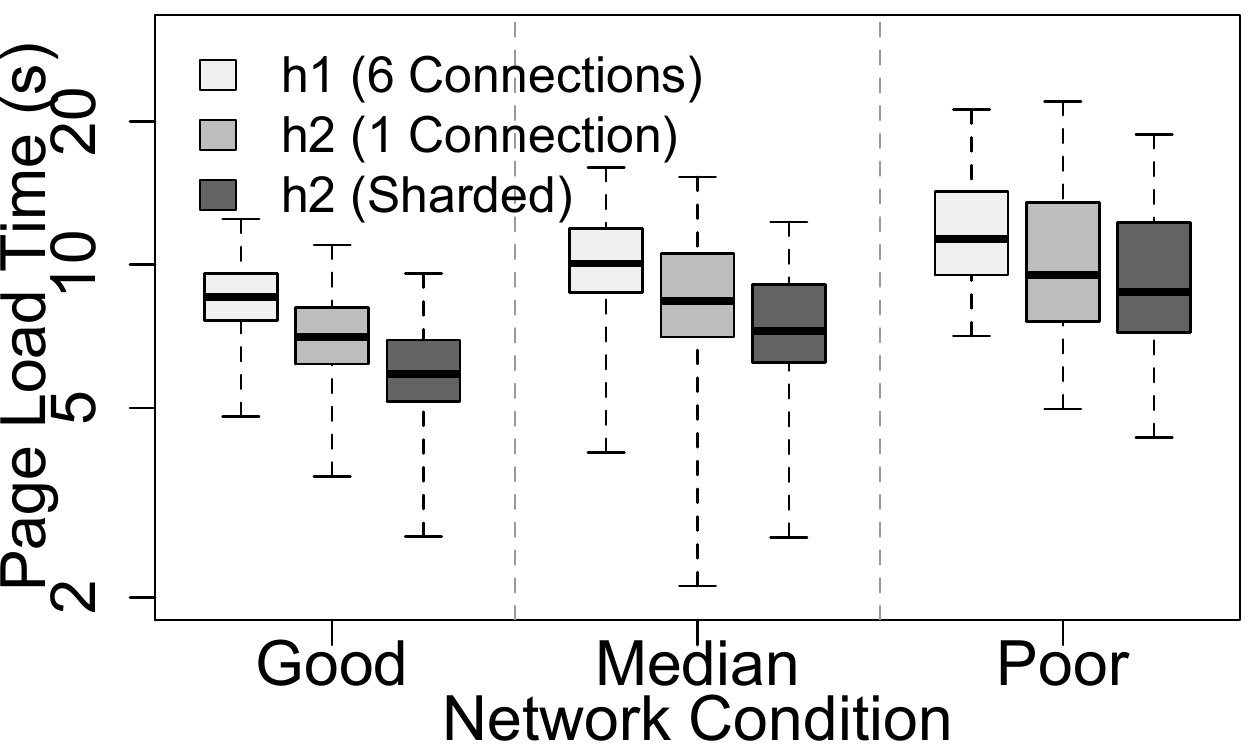}
\vspace{-15pt}
\caption{PLT distributions for a 2\,MB page when emulating based on TCP loss frequency.}
\label{fig:2mb_new_data}
\endminipage
\vspace{-10pt}
\end{figure*}

\begin{figure*}[t]
\centering
\minipage{0.325\textwidth}
  \includegraphics[width=\linewidth]{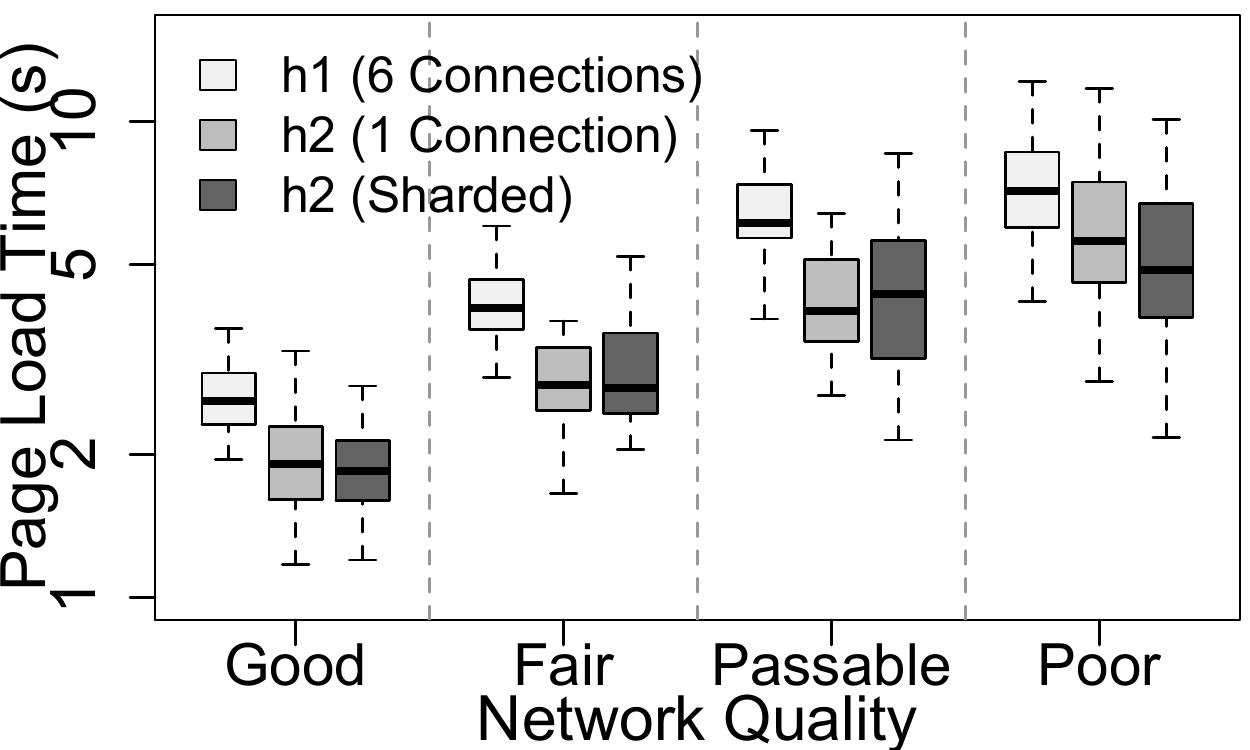}
\vspace{-15pt}
\caption{PLTs of a 2\,MB page when emulating based on network quality.}
\label{fig:2mb_poster}
\endminipage
\hfill
\minipage{0.325\textwidth}
  \includegraphics[width=\linewidth]{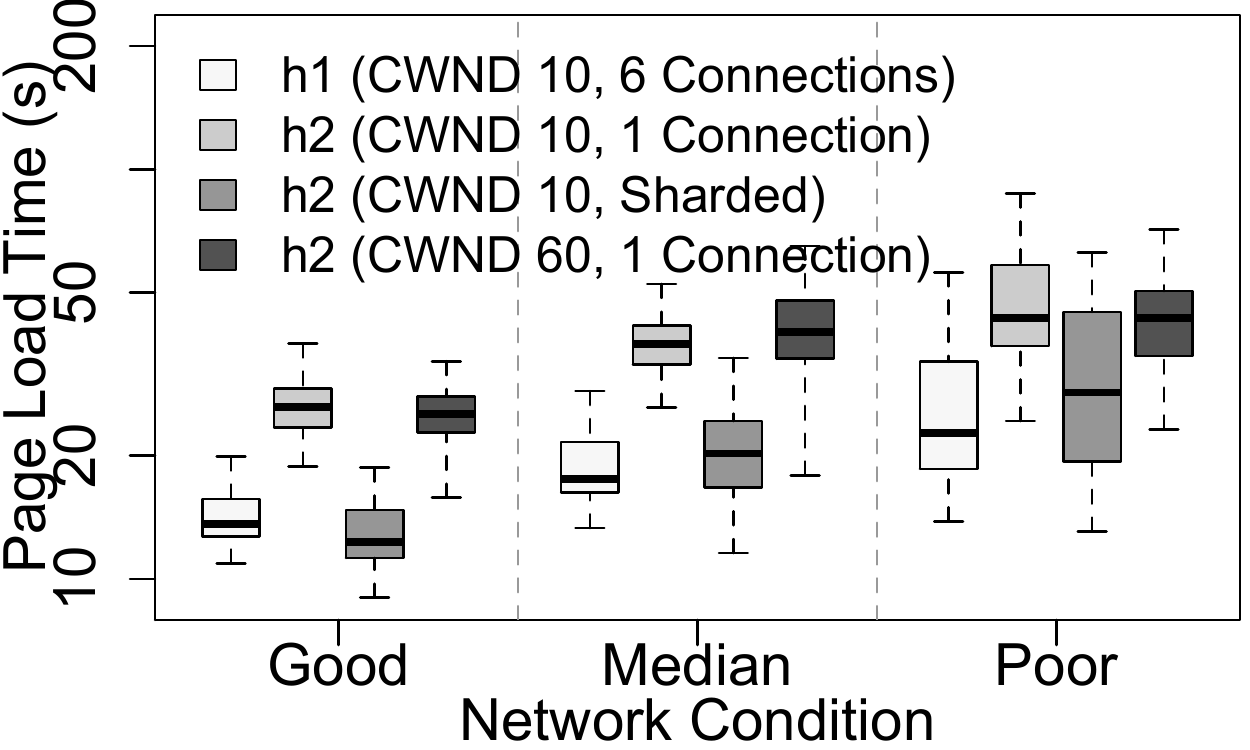}
\vspace{-15pt}
\caption{PLTs of 8\,MB page when emulating network based on TCP loss frequency.}
\label{fig:8mb_new_data}
\endminipage
\hfill
\minipage{0.325\textwidth}
  \includegraphics[width=\linewidth]{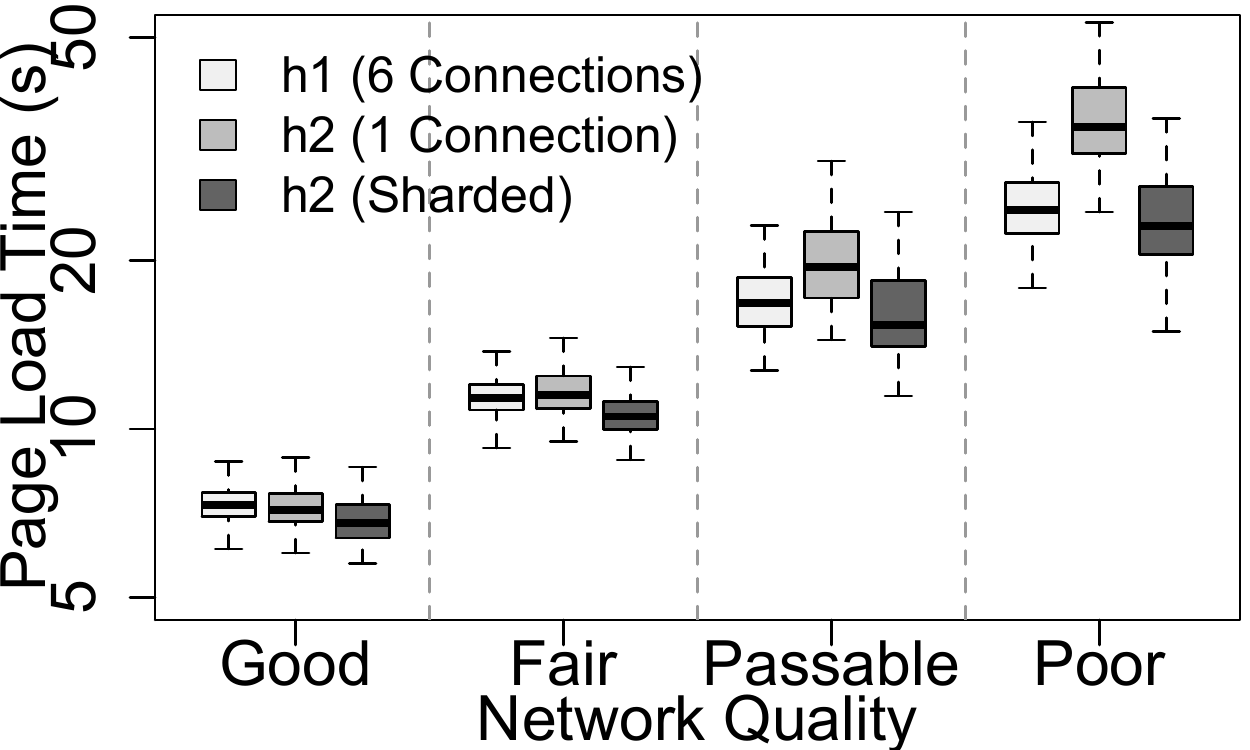}
\vspace{-15pt}
\caption{PLTs of a 8\,MB page when emulating based on network quality.}
\label{fig:8mb_poster}
\endminipage
\vspace{-15pt}
\end{figure*}

When using the emulation scenarios described in Section~\ref{sec:condition}, we found that the results were qualitatively similar to what we describe above.
Therefore, due to the page limit, we do not show or discuss those results explicitly in the paper, however we briefly discuss them using Figures~\ref{fig:spinning_shard},~\ref{fig:moritz_shard},~\ref{fig:2mb_new_data}, and~\ref{fig:8mb_new_data} in Section~\ref{sec:sharding}.
\looseness -1

\vspace{-5pt}
\section{\texttt{h2} performance with Domain-sharding}
\label{sec:sharding}

From the previous section we observe that in the event of packet loss, \texttt{h2} degrades PLTs that contain both large and small objects multiplexed over the same TCP connection.
In this section, we investigate whether PLTs over \texttt{h2} can be improved by isolating large and small object transfers on different connections, such that their downloads do not interfere, especially during loss.
We speculate that isolating large and small objects onto separate connections should allow the server to transfer all small objects in fewer round trips using \texttt{h2}'s multiplexing~(similarly to Figure~\ref{fig:spinning}, and as well as prevent the downloads of large objects from blocking smaller objects in the event of loss~(similarly to Figure~\ref{fig:javier}).
However, unlike \texttt{h1}, modern Web browsers, such as Google Chrome and Mozilla Firefox, establish only one TCP connection for every \texttt{h2}-compatible domain name on the webpage.
To the best of our knowledge, there is currently no provision in the browser source code to allow establishment of multiple \texttt{h2} connections, without modifying the underlying \texttt{h2} protocol~\cite{quic,smig}.
Besides increasing the number of TCP connections used for \texttt{h2} as we do in this paper, one could also tune TCP specifically for \texttt{h2}, for example, by increasing the ICW to 6 times the size of what is recommended for \texttt{h1},~i.e.~60 segments.
However, we will discuss the performance of this approach later in the section.
\looseness -1

We leverage domain-sharding by using multiple \texttt{h2}-compatible domain names on webpages to enable the browser to establish multiple \texttt{h2} connections.
Specifically, we setup multiple \texttt{h2}-compatible domain names on the \texttt{server} to shard webpage objects on these domain names. 
Note that we associate each domain name with a unique certificate to avoid connection coalescing used by modern Web browsers.
\looseness -1

We then investigate how webpage objects should be sharded such that the impact of packet loss on \texttt{h2} is minimized.
For this investigation, we use the 8\,MB page~(with 12 large objects and 124 small objects) used for Figure~\ref{fig:javier} and create several versions of this page. 
Each version contains some number of large objects isolated on different connections via domain-sharding.
Specifically, in Figure~\ref{fig:sharding_effect}, \textit{Sharding Type A} refers to the page where we isolate only two large objects on different connections, and all other objects on one connection -- total of three connections.
\textit{Sharding Type B} refers to the page where we isolate all 12 large objects on different connections and all small objects on one connection -- total of 13 connections.
\textit{Sharding Type C} refers to the page where we isolate five large objects on different connections, and all other objects on one connection -- total of six connections.
The page loads labeled as \texttt{h2} show PLTs using the original 8\,MB page over single connection.
\looseness -1

From Figure~\ref{fig:sharding_effect}, we observe that PLTs for sharded webpages are always lower than PLTs over \texttt{h2} with one connection.
Moreover, \textit{Sharding Type B} offers the lowest PLTs among all sharded webpages, as this approach speeds up transfers of small objects and prevents large objects to impact small object downloads during loss.
Therefore, isolating each large download on a separate connection is a reasonable strategy to reduce the PLT of \texttt{h2}-enabled webpages. 
\looseness -1

\vspace{-4pt}
\subsection{Measuring Web performance with Sharded-\texttt{h2}}

In Figure~\ref{fig:spinning_shard}, we show the distributions of PLTs for a webpage with 365 objects of size 1\,KB each loaded over \texttt{h2} with one connection, \texttt{h1} with six connections, and \texttt{h2} with multiple connections, using the emulation scenarios described in Section~\ref{sec:condition}.
Similar to Figure~\ref{fig:spinning}, we observe that \texttt{h2} with one connection outperforms \texttt{h1} with six connections.
When sharding half of the objects on a different \texttt{h2}-compatible domain name, i.e. using two connections, we observe that in \textit{Good} conditions multiple \texttt{h2} connections improve the PLTs by 12\% in the median case.
Further, when sharding the page with three \texttt{h2}-compatible domain names, i.e. using three connections, we observe that the PLTs are either comparable or slightly higher than PLTs over \texttt{h2} with two connections.
For webpages with hundreds of small-sized objects, establishing multiple connections introduce additional latency of TCP and TLS handshakes to the overall PLT -- negating the benefits of domain-sharding.
\looseness -1

Next, in Figure~\ref{fig:moritz_shard} we show PLT distributions of a webpage with 10 large objects of size 435\,KB each, using the emulation scenarios described in Section~\ref{sec:condition}.
For this experiment, we create multiple copies of the webpage and shard the resources on different \texttt{h2}-compatible domain names, such that the browser establishes two, three, six, and 10 \texttt{h2} connections depending upon how many domain names we use on the webpage.
Similarly to Figure~\ref{fig:moritz}, we observe that \texttt{h2} with one connection degrades PLT when compared to \texttt{h1} with six connections.
We also observe that as the number of \texttt{h2} connections increase, the PLTs decrease under all network conditions. 
For example, under \textit{Good} network conditions, the median PLT over \texttt{h1} with six connections is 8\,seconds, \texttt{h2} with one connection is 15\,seconds, and PLTs over \texttt{h2} with two, three, six, 10 connections are 12, 10, 9, and 8\,seconds respectively.
Although, for such a webpage multiple \texttt{h2} connections do not result in PLTs lower than \texttt{h1} but only comparable, we argue that loading the webpage with   sharding provides significantly lower PLTs compared to webpages loaded over \texttt{h2} with one connection.
We also measure the PLT over \texttt{h2} when using one connection with a ICW of 60 segments.
Similar to Figure~\ref{fig:8mb_new_data}, we observe that a high ICW does not improve PLTs, because in lossy network conditions the rate at which the congestion window re-grows after loss remains same as the rate it grows when using one connection with ICW of 10.
\looseness -1

In Figure~\ref{fig:2mb_new_data}, we show the PLT distribution of a 2\,MB webpage~(same page as used in Figure~\ref{fig:javier}) when loaded over \texttt{h2} with one connection, \texttt{h1} with six connections, and \texttt{h2} with four connections~(one for all small objects and the other three for the three large objects).
Similarly to Figure~\ref{fig:javier}, in Figure~\ref{fig:2mb_new_data} we observe that \texttt{h2} outperforms \texttt{h1} in terms of PLT.
Moreover, \texttt{h2} with four connections achieves even faster PLTs, compared to both \texttt{h2} with one connection and \texttt{h1} with six connections.
\looseness -1

When emulating the network based on network quality as shown in Figure~\ref{fig:2mb_poster}, we observe that in the case of \textit{Good}, \textit{Fair}, and \textit{Passable} user experiences, sharded-\texttt{h2} with four connections achieves either 3\% faster or comparable PLTs to \texttt{h2} with one connection.
However, in the case of \textit{Poor} user experience, sharded-\texttt{h2} offers 13\% faster PLTs than \texttt{h2} with one connection.
\looseness -1

Next, in Figure~\ref{fig:8mb_new_data}, we show PLT distributions of a 8\,MB webpage, when loaded using the emulation scenarios described in Section~\ref{sec:condition}.
For this experiment, we used one domain name for all small objects and 12 different domain names for the 12 large objects.
From the figure we observe that under \textit{Good} network conditions, sharded-\texttt{h2} offers 53\% faster PLTs compared to \texttt{h2} with one connection and 10\% faster PLT compared to \texttt{h1} with six connections.
Even when the network conditions get worse, sharded-\texttt{h2} outperforms \texttt{h2} with one connection, however, in such conditions sharded-\texttt{h2} yields PLTs higher or comparable to \texttt{h1}.
Finally, and similarly to Figure~\ref{fig:moritz_shard}, we observe that PLTs over \texttt{h2} with one connection using ICW of 60 are comparable to PLTs using one connection with ICW of 10.
\looseness -1

Our results for sharded-\texttt{h2} for the 12\,MB page in both emulation setups are qualitatively similar to the ones shown in Figures~\ref{fig:8mb_new_data} and~\ref{fig:8mb_poster}.
\looseness -1

\noindent
\textit{Note:} Instead of using 12 \texttt{h2}-compatible hostnames, one could use two \texttt{h1} hostnames, for each of which browsers establish six connections.
Also, as browsers can not infer object sizes from the base page HTML, our sharding technique enables CPs to instruct browsers on how to fetch objects on different connections, based on their sizes, for improved performance.
\looseness -1

\vspace{-5pt}
\section{Challenges in Validating the Emulator}
\label{sec:validation}

As mentioned in Section~\ref{sec:why_testbed}, RUM-based systems that leverage browser provided APIs, such as the Navigation Timing API~\cite{navigationTimingApi}, do not capture TCP metrics and thus their estimation of PLT does not indicate whether the page load experienced loss, and if so, how much~\cite{akamai:rum}.
While CDN providers capture TCP metrics pertaining to the recorded RUM data, 
both RUM and TCP-based metrics are independently recorded at a very low sample rate of 1-5\%.
Despite this fact, we sought to find page load sessions for which both RUM and TCP-based metrics were recorded, but because of the low sampling rate the common set was too small to perform meaningful analysis~\cite{goelAkamaiSlides16}.
\looseness -1

Additionally, since about 32\% of the TCP connections experience loss, RUM-based PLT estimations include page loads from sessions with and without loss.
On the other hand, we emulate only lossy network conditions.
And so, the comparison of PLT distributions across the two techniques does not help validate our mobile emulator.
\looseness -1

Next, our data collection technique relies on capturing TCP traffic using \texttt{TCPDump}, which do not indicate the deployed cellular network technology~(2G/3G/LTE). 
Therefore, we can not comment on the degree of fidelity to which our cellular emulator can emulate radio access networks.
However, we speculate that our classification of captured TCP connections into various network qualities could potentially reflect on the different cellular technologies deployed by the carrier.
For example, the \textit{Good}, \textit{Median}, and \textit{Poor} network conditions could potentially represent the behavior of transport protocol over LTE, 3G, and 2G configurations respectively.
\looseness -1

In summary, the current Web performance measurement techniques do not allow us to validate the emulator we developed -- motivating further research into cellular network measurement and emulation.
However, we believe that seeding the emulator with real cellular network data enables us to emulate the dynamic characteristics of cellular networks, with sufficient fidelity to reason about their impact on \texttt{h2} performance.
\looseness -1

\vspace{-8pt}
\section{Discussion on Domain-Sharding}
\label{sec:discussion}

Since our internal investigation of loss~(not shown) indicates little-to-no loss in wired access networks and since our study only focuses on cellular network conditions, we recommend the use of domain-sharding for webpages served to mobile clients only.
However, depending on how domain-sharding is employed by Web developers and CDN providers, its use may pose several potential implications on Web performance. 
\looseness -1

The use of domain-sharding incurs additional DNS lookups, which may take several hundred milliseconds in cellular networks~\cite{GoelIPv6,GoelMiddlebox16,Rula:2014:BCC:2663716.2663734}, and can potentially increase the overall PLT.
However, several Web optimization techniques, such as DNS Pre-Resolve~\cite{msdn:dns-prefetch} and DNS PiggyBacking~\cite{ShangPiggy06}, can eliminate this additional lookup latency by providing hints in the base page HTML, such that the Web browser resolves the domain names much before the resolution is needed.
Other hints, such as TCP Pre-Connect, can also help browsers help establish TCP connections much before they are needed~\cite{tcpPreConnect}.
\looseness -1

Additional connections via domain-sharding also requires additional computational resources for cryptographic operations~\cite{costOfS}.
However, major CDN providers today use the latest cipher suites, such as ChaCha20 and Poly1305, that offer improved mobile Web performance, even when compared to the unencrypted Web~\mbox{\cite{chachapoly:rfc,chachapoly}}.
Moreover, for many years domain-sharding has been in use with \texttt{h1}, and therefore the needed resources are already available to support even up to six encrypted connections for each domain name.
We argue that using domain-sharding with \texttt{h2} will require far fewer secure connections, as unlike \texttt{h1}, a single \texttt{h2} connection can be used to download all small objects -- only leaving a few big objects to be downloaded on separate connections. 
\looseness -1

\vspace{4pt}
\noindent
\textit{Guidelines to applying domain-sharding in practice:} Many CDN providers use several front end optimization~(FEO) techniques to generate the base page HTML based on the type of client device~(mobile vs desktop), client's Web browser, device screen size, client's ISP, last-mile performance, among many other factors~\cite{imageManager}.
As such, our proposed domain-sharding can serve as another optimization technique that FEO engines could incorporate to generate different versions of webpages -- pages with and without domain-sharding.
Similarly to the existing FEO implementations, the different versions of the same webpage could be cached by CDN servers and the sharded version could be served to cellular clients.
\looseness -1

\vspace{-8pt}
\section{Conclusions}
\label{sec:conclusions}
\vspace{-3pt}

\texttt{h2} eliminates HOL blocking at the application layer but retains it on the transport layer, which impacts Web performance in lossy cellular network conditions. 
In this work, we study the characteristics of TCP connections observed for clients in a major US cellular carrier.
We then model various cellular network conditions to investigate \texttt{h2} performance under lossy network conditions. 
Our results indicate that \texttt{h2} offers faster PLTs when webpages contain small objects, however, \texttt{h2} degrades PLTs when downloading webpages with many large objects.
Using domain-sharding, we demonstrate that loading webpages over \texttt{h2} with multiple TCP connections reduces the impact of packet loss on PLT -- improving Web performance for mobile clients.
Based on our experimental evaluation of applying domain-sharding on \texttt{h2}-enabled webpages, we recommend CPs and CDN providers to apply \texttt{h2}-aware domain-sharding practices when upgrading mobile Web content delivery to \texttt{h2} protocol.
\looseness -1

\vspace{4pt}
\noindent
\textbf{ACKNOWLEDGMENTS:} We thank Javier Garza, Ajay Kumar, and Kanika Shah for their help in setting experiments and refining our research directions.
We also thank NSF for supporting us via grants CNS-1555591 and CNS-1527097.
\looseness -1

\vspace{-8pt}
\bibliographystyle{abbrv}
\small
\bibliography{h2}

\begin{thebibliography}{10}

\bibitem{webpagetest}
{Test a Website's Performance}.
\newblock \url{http://www.webpagetest.org/}, 2008.

\bibitem{gomez}
{Gomez Last-Mile Testbed}.
\newblock
  \url{http://www.sqaforums.com/attachments/601980-SQA_Gomez_DollarThrifty_Webinar_QandA.PDF},
  Nov. 2009.

\bibitem{mobitest}
{Akamai Mobitest}.
\newblock \url{http://mobitest.akamai.com/m/index.cgi}, 2012.

\bibitem{4GMark}
{4GMark: Mobile Performance test}.
\newblock \url{http://www.4gmark.com/}, Nov. 2014.

\bibitem{telemetry}
{Chromium Telemetry}.
\newblock \url{https://www.chromium.org/developers/telemetry/run_locally}, Dec.
  2014.

\bibitem{nperf}
{What’s nPerf? How does it work?}
\newblock \url{http://www.nperf.com/en/}, Nov. 2014.

\bibitem{disableSharding1}
{7 Tips for Faster HTTP/2 Performance}.
\newblock
  \url{https://www.nginx.com/blog/7-tips-for-faster-http2-performance/}, Oct.
  2015.

\bibitem{akamai:rum}
{Akamai Real User Monitoring}.
\newblock
  \url{https://www.akamai.com/us/en/resources/real-user-monitoring.jsp}, Aug.
  2015.

\bibitem{navigationTimingApi}
{Navigation Timing}.
\newblock \url{http://w3c.github.io/navigation-timing/}, Aug. 2015.

\bibitem{radioopt}
{RadioOpt}.
\newblock \url{https://www.radioopt.com/}, Mar. 2015.

\bibitem{aanp}
{Akamai Accelerated Network Partner}.
\newblock
  \url{https://www.akamai.com/us/en/multimedia/documents/akamai/akamai-accelerated-network-partner-aanp-faq.pdf},
  Jun. 2016.

\bibitem{imageManager}
{Akamai Image Manager}.
\newblock
  \url{https://www.akamai.com/us/en/products/web-performance/image-manager.jsp},
  Jun. 2016.

\bibitem{keynote}
{Dynatrace Synthetic Monitoring}.
\newblock
  \url{http://www.keynote.com/solutions/monitoring/dynatrace-synthetic-monitoring},
  2016.

\bibitem{httparchiveDownloads}
{HTTP Archive: Downloads}.
\newblock \url{http://mobile.httparchive.org/downloads.php}, Apr. 2016.

\bibitem{httparchive}
{HTTP Archive: Interesting stats}.
\newblock
  \url{http://mobile.httparchive.org/interesting.php?a=All&l=Apr%201%202016&s=Top1000},
  Apr. 2016.

\bibitem{tc}
{Netem}.
\newblock \url{https://wiki.linuxfoundation.org/networking/netem}, Jul. 2016.

\bibitem{tshark}
{Tshark}.
\newblock \url{https://www.wireshark.org/docs/man-pages/tshark.html}, 2016.

\bibitem{rfc2581}
M.~Allman, V.~Paxson, and W.~Stevens.
\newblock {TCP Congestion Control, RFC 2581}, Apr. 1999.

\bibitem{http2:rfc}
M.~Belshe, R.~Peon, and E.~M.~Thomson.
\newblock {Hypertext Transfer Protocol Version 2 (HTTP/2), RFC 7540}, May 2015.

\bibitem{Bischof2017}
Z.~S. Bischof, F.~E. Bustamante, and R.~Stanojevic.
\newblock {\em The Utility Argument -- Making a Case for Broadband SLAs}.
\newblock Mar. 2017.

\bibitem{Bocchi2017}
E.~Bocchi, L.~De~Cicco, M.~Mellia, and D.~Rossi.
\newblock {\em The Web, the Users, and the MOS: Influence of HTTP/2 on User
  Experience}.
\newblock 2017.

\bibitem{bbr}
N.~Cardwell, Y.~Cheng, C.~S. Gunn, S.~H. Yeganeh, and V.~Jacobson.
\newblock {BBR: Congestion-Based Congestion Control}.
\newblock {\em ACM Queue}, Sept. 2016.

\bibitem{7179400}
H.~de~Saxcé, I.~Oprescu, and Y.~Chen.
\newblock {Is HTTP/2 really faster than HTTP/1.1?}
\newblock In {\em IEEE INFOCOM WKSHPS}, April 2015.

\bibitem{msdn:dns-prefetch}
G.~Developers.
\newblock {Pre-Resolve DNS}.
\newblock
  \url{https://developers.google.com/speed/pagespeed/service/PreResolveDns},
  Apr. 2015.

\bibitem{Erman:2013:TSM:2535372.2535399}
J.~Erman, V.~Gopalakrishnan, R.~Jana, and K.~K. Ramakrishnan.
\newblock {Towards a SPDY'ier Mobile Web?}
\newblock In {\em ACM CoNEXT}, Dec. 2013.

\bibitem{45411}
T.~Flach, P.~Papageorge, A.~Terzis, L.~Pedrosa, Y.~Cheng, T.~Karim,
  E.~Katz-Bassett, and R.~Govindan.
\newblock {An Internet-Wide Analysis of Traffic Policing}.
\newblock In {\em ACM SIGCOMM}, Aug. 2016.

\bibitem{rfc5690}
S.~Floyd, A.~Arcia, D.~Ros, and J.~Iyengar.
\newblock {Adding Acknowledgement Congestion Control to TCP, RFC 5690}, Feb.
  2010.

\bibitem{goelAkamaiSlides16}
U.~Goel.
\newblock {Web Performance in Cellular Networks}.
\newblock
  \url{http://www.cs.montana.edu/~utkarsh.goel/docs/Goel_Akamai_Intern_Showcase_2016.pdf},
  Jul. 2016.

\bibitem{GoelIPv6}
U.~Goel, M.~Steiner, M.~P. Wittie, M.~Flack, and S.~Ludin.
\newblock {A Case for Faster Mobile Web in Cellular IPv6 Networks}.
\newblock In {\em MobiCom}, Oct. 2016.

\bibitem{GoelMiddlebox16}
U.~Goel, M.~Steiner, M.~P. Wittie, M.~Flack, and S.~Ludin.
\newblock {Detecting Cellular Middle-boxes using Passive Measurement
  Techniques}.
\newblock In {\em ACM PAM}, Mar. 2016.

\bibitem{GoelH2MobiCom16}
U.~Goel, M.~Steiner, M.~P. Wittie, M.~Flack, and S.~Ludin.
\newblock {HTTP/2 Performance in Cellular Networks}.
\newblock In {\em ACM MobiCom}, Oct. 2016.

\bibitem{GoelThird}
U.~Goel, M.~Steiner, M.~P. Wittie, M.~Flack, and S.~Ludin.
\newblock {Measuring What is Not Ours: A Tale of 3rd Party Performance}.
\newblock In {\em ACM PAM}, Mar. 2017.

\bibitem{GoelSurveyE2E}
U.~Goel, M.~P. Wittie, K.~C. Claffy, and A.~Le.
\newblock {Survey of End-to-End Mobile Network Measurement Testbeds, Tools, and
  Services}.
\newblock {\em IEEE Communications Surveys Tutorials}, 18(1), Firstquarter
  2016.

\bibitem{7288411}
U.~Goel, M.~P. Wittie, and M.~Steiner.
\newblock {Faster Web through Client-Assisted CDN Server Selection}.
\newblock In {\em IEEE Conference on Computer Communication and Networks
  (ICCCN)}, Aug 2015.

\bibitem{grigorik2013high}
I.~Grigorik.
\newblock {\em {High Performance Browser Networking}}.
\newblock {O'Reilly Media, Inc.}, Sept. 2013.

\bibitem{disableSharding2}
I.~Grigorik.
\newblock {Making the Web Faster with HTTP 2.0}.
\newblock {\em ACM Queue}, 11(10), Oct. 2013.

\bibitem{tcpPreConnect}
I.~Grigorik.
\newblock {Eliminating Roundtrips with Preconnect}.
\newblock
  \url{https://www.igvita.com/2015/08/17/eliminating-roundtrips-with-preconnect/},
  Aug. 2015.

\bibitem{cubic}
S.~Ha, I.~Rhee, and L.~Xu.
\newblock {CUBIC: A New TCP-friendly High-speed TCP Variant}.
\newblock {\em SIGOPS Operating System Review}, 42(5), July 2008.

\bibitem{quic}
R.~Hamilton, J.~Iyengar, I.~Swett, and A.~Wilk.
\newblock {QUIC: A UDP-Based Secure and Reliable Transport for HTTP/2, Draft},
  Jan. 2016.

\bibitem{disableSharding3}
R.~Hodson.
\newblock {HTTP/2 For Web Developers}.
\newblock \url{https://blog.cloudflare.com/http-2-for-web-developers/}, Dec.
  2015.

\bibitem{7557456}
Y.~Liu, Y.~Ma, X.~Liu, and G.~Huang.
\newblock {Can HTTP/2 Really Help Web Performance on Smartphones?}
\newblock In {\em IEEE SCC}, Jun. 2016.

\bibitem{ludinH2}
S.~Ludin and J.~Garza.
\newblock {\em {Learning HTTP/2: An Introduction to the Next Generation Web}}.
\newblock {O'Reilly Media, Inc.}, Dec. 2016.

\bibitem{smig}
X.~Mi, F.~Qian, and X.~Wang.
\newblock {SMig: Stream Migration Extension for HTTP/2}.
\newblock In {\em ACM CoNEXT}, Dec. 2016.

\bibitem{costOfS}
D.~Naylor, A.~Finamore, I.~Leontiadis, Y.~Grunenberger, M.~Mellia,
  M.~Munaf\`{o}, K.~Papagiannaki, and P.~Steenkiste.
\newblock {The Cost of the "S" in HTTPS}.
\newblock In {\em ACM CoNEXT}, Dec. 2014.

\bibitem{chachapoly:rfc}
Y.~Nir and A.~Langley.
\newblock {ChaCha20 and Poly1305 for IETF Protocols, RFC 7539}, May 2015.

\bibitem{disableSharding4}
G.~Polvara.
\newblock {Mind the Gap When Upgrading to HTTP/2}.
\newblock
  \url{http://nerds.fundbase.com/2015/08/20/mind-the-gap-when-upgrading-to-http-2/},
  Aug. 2015.

\bibitem{Rula:2014:BCC:2663716.2663734}
J.~P. Rula and F.~E. Bustamante.
\newblock {Behind the Curtain: Cellular DNS and Content Replica Selection}.
\newblock In {\em ACM IMC}, Nov. 2014.

\bibitem{ShangPiggy06}
H.~Shang and C.~E. Wills.
\newblock {Piggybacking Related Domain Names to Improve DNS Performance}.
\newblock {\em Computer Network}, 50(11), Aug. 2006.

\bibitem{moritzHttpWorkshop16}
M.~Steiner.
\newblock {H2 Performance Analysis in Real World Cellular Networks}.
\newblock
  \url{https://www.ietf.org/proceedings/97/slides/slides-97-maprg-h2-performance-analysis-in-real-world-cellular-networks-moritz-steiner-01.pdf},
  Jul. 2016.

\bibitem{DBLP:journals/corr/SteinerG16}
M.~Steiner and R.~Gao.
\newblock {What slows you down? Your network or your device?}
\newblock {\em CoRR}, abs/1603.02293, 2016.

\bibitem{chachapoly}
N.~Sullivan.
\newblock {Do the ChaCha: Better Mobile Performance with Cryptography}.
\newblock
  \url{https://blog.cloudflare.com/do-the-chacha-better-mobile-performance-with-cryptography/},
  Feb. 2015.

\bibitem{linkConditioner}
M.~Thompson.
\newblock {Network Link Conditioner}.
\newblock \url{http://nshipster.com/network-link-conditioner/}, Sept. 2013.

\bibitem{pamhttp2}
M.~Varvello, K.~Schomp, D.~Naylor, J.~Blackburn, Alessandro, Finamore, and
  K.~Papagiannaki.
\newblock {Is The Web HTTP/2 Yet?}
\newblock In {\em PAM}, Mar. 2016.

\bibitem{wprof}
X.~S. Wang, A.~Balasubramanian, A.~Krishnamurthy, and D.~Wetherall.
\newblock {Demystifying Page Load Performance with WProf}.
\newblock In {\em USENIX NSDI}, Apr., 2013.

\bibitem{Wang:2014:SS:2616448.2616484}
X.~S. Wang, A.~Balasubramanian, A.~Krishnamurthy, and D.~Wetherall.
\newblock {How Speedy is SPDY?}
\newblock In {\em USENIX NSDI}, Apr. 2014.

\bibitem{sprout}
K.~Winstein, A.~Sivaraman, and H.~Balakrishnan.
\newblock {Stochastic Forecasts Achieve High Throughput and Low Delay over
  Cellular Networks}.
\newblock In {\em USENIX NSDI}, Apr. 2013.

\bibitem{Xu:2011:CDN:1993744.1993777}
Q.~Xu, J.~Huang, Z.~Wang, F.~Qian, A.~Gerber, and Z.~M. Mao.
\newblock {Cellular Data Network Infrastructure Characterization and
  Implication on Mobile Content Placement}.
\newblock In {\em ACM SIGMETRICS}, 2011.

\bibitem{ethanCellProxy}
X.~Xu, Y.~Jiang, T.~Flach, E.~Katz-Bassett, D.~Choffnes, and R.~Govindan.
\newblock {Investigating Transparent Web Proxies in Cellular Networks}.
\newblock In {\em ACM PAM}, Mar. 2015.

\bibitem{Zarifis2014}
K.~Zarifis, T.~Flach, S.~Nori, D.~Choffnes, R.~Govindan, E.~Katz-Bassett, Z.~M.
  Mao, and M.~Welsh.
\newblock {Diagnosing Path Inflation of Mobile Client Traffic}.
\newblock In {\em ACM PAM}, 2014.

\end{thebibliography}
\end{document}